# 北京邮電大學

## 硕士学位论文

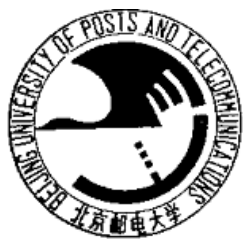

题目： __基于 SDR 的 LTE-WiFi 聚合系统的__
　　　　　__设计与实现__

学　　号：　__2015110336__
姓　　名：　__刘海涛__
专　　业：　__电子与通信工程__
导　　师：　__王健全__
学　　院：　__信息与通信工程__

**2017 年 11 月 29 日**

# 基于 SDR 的 LTE-WiFi 聚合系统的设计与实现

## 摘　要

近年来，移动互联网接入数据呈现出爆炸式增长，给运营商带来了巨大的流量负载。而 WiFi 分流作为一种重要的无线通信技术，在减轻移动网络压力和提高网络服务方面拥有巨大的潜力。在此基础上，LTE-WiFi 聚合 (LTE-WiFi Aggregation，LWA)技术实现了从蜂窝网络转移流量到 WiFi 网络，充分利用了传输资源，是一种非常有发展前景的技术。而软件无线电 (Software Defined Radio, SDR)技术，通过软件的更新升级取代硬件研究的高消耗，为通信技术的验证和实现提供了扎实、便捷的技术平台。因此，研究和设计基于 SDR 的 LWA 系统对于利用 SDR 和验证通信技术都有重大意义。

本文绪论部分主要介绍了 LWA 技术的研究背景和研究现状，在第二部分对 WiFi 融合技术和软件无线电技术做了简单的总结。首先介绍了 WiFi 融合技术的发展历程，通过比较对 LWA 技术进行了系统分析。之后对软硬件平台进行介绍，包括由于实现完整的通信协议被广泛关注的软件平台 OpenAirInterface(OAI) 和硬件设备通用软件无线电外设(Universal Software Radio Peripheral, USRP)。

第三部分主要介绍了基于 SDR 的 LTE-WiFi 聚合系统的设计与实现。首先，简要介绍了 LTE-WiFi 聚合技术原理和开发环境的搭建，从原理上对 LWA 基站和用户侧的设计进行了分析并据此对原有的平台做出改善。然后分别详细的阐述了 LWA 基站侧和用户侧的设计与实现。首先给出了 LWA 聚合系统的工作流程的设计，然后深入分析了 OAI 协议栈优化，分流接口和用户侧重排序等模块。最后，根据实际空口环境下的实验结果，验证了本文设计实现的 LTE-WiFi 聚合系统的正确性及 LTE-WiFi 聚合技术的性能优势。

第四部分则是基于 LTE-WiFi 聚合系统提出了一种分流策略算法。第三部分系统着重于实现 WiFi 分流的系统整体架构实现。实际上，基站侧应该根据实时的网络状况对 LTE 和 WiFi 网络资源进行快速、动态的调整，以最大化的利用资源。这里提出一种算法，建立数据缓存，统计链路负载，然后通过判断根据两条链路的网络负载情况，实时地修改分流比例，做到动态调整。通过空口实验验证了分流策略的正确性，证明优化后的系统有

更好的处理效果，并且仍然有很大的提升空间，能够为以后验证通信新技术提供可靠的参考。

# DESIGN AND IMPLEMENTATION OF A LTE-WIFI AGGREGATION SYSTEM BASED ON SDR

## ABSTRACT

In recent years, the mobile Internet access data show an explosive growth, bringing huge traffic load to operators. WiFi offloading, as an important wireless communication technology, has great potential to alleviate the pressure of mobile network and improve network services. On this basis, LTE-WiFi Aggregation (LTE-WiFi Aggregation, LWA) technology is a very promising technology for transferring traffic from cellular network to WiFi network and making full use of transmission resources. Software Defined Radio (SDR) technology replaces the high consumption of hardware research by updating and upgrading software, and provides a solid and convenient technology platform for the verification and implementation of communication technologies. Therefore, studying and designing a SDR-based LWA system is of great importance for both utilizing SDR technologie and verifying communication technologies.

The introductory part of this article mainly introduces the research background and status of LWA technology. In the second part, we briefly summarize the WiFi fusion technology and software radio technology. First, the development of WiFi fusion technology is introduced. The analysis of LWA technology is specific by comparing. The hardware and software platforms introductions are followed, including the OpenAirInterface (OAI) software platform and the Universal Software Radio Peripheral (USRP), a hardware device widely watched for implementing complete communication protocols.

The third part mainly introduces the design and implementation of LTE-WiFi aggregation system based on SDR. First of all, it briefly introduced the construction of LTE-WiFi aggregation technology and development environment, and analyzed the design of LWA base station and user side in principle and improved the original platform accordingly. Then elaborated the design and implementation of LWA base station side and user side respectively. At first, the design of the workflow of LWA aggregation system is given, and then the modules of OAI protocol stack optimization, split interface and user-oriented sorting are deeply analyzed. Finally, we verify the correctness of the LTE-WiFi aggregation system designed and implemented in this paper and the performance advantages of the LTE-WiFi aggregation technology according to the experimental results under the actual air interface environment.

The fourth part is based on LTE-WiFi aggregation system proposed an offloading strategy algorithm. The third part of the system focuses on the realization of the overall architecture of the system to achieve WiFi offload. In fact, the base station side should make rapid and dynamic adjustments to the LTE and WiFi network resources according to the real-time network conditions in order to maximize the use of resources. Here an algorithm is proposed to set up a data cache and count the load of the link. Then, by judging the load of the network according to the two links, the proportion of shunts can be modified in real time so as to be dynamically adjusted. The correctness of the offloading strategy is verified through the air interface experiment, which proves that the optimized system has a better processing effect and still has a lot of space for improvement. It can provide a reliable reference for the future verification of new communication technologies.

# 目　录

# 第一章 绪论

## 1.1 研究背景

近年来，互联网的迅猛发展为移动互联网、无线通信等应用提供了有力的技术支持，伴随着移动网络(如 3G/LTE/4G)、移动无线局域网通信技术(如 WiFi/Bluetooth/NFC/D2D)的推广和普及，拥有无线通讯能力的智能设备越来越多，随之而来的是移动数据的爆炸式增长。根据互联网的估计，全球移动数据流量预计在 2015 年至 2020 年间将增长近十倍。在 2020 年，全球移动数据流量预计将达到每月 30.6 亿字节，复合年增长率为 53％[1]。 尽管近年来 LTE 技术有所发展和普及，但是运营商网络仍然难以满足移动设备的需求。日益增长的移动数据造成的巨大的流量负载和网络拥塞，逐渐成为移动宽带时代发展的"瓶颈"。

无线局域网是计算机网络与无线通信技术相结合的产物，WiFi 通过电磁波发送和接收数据，可以提供有线局域网的所有功能[2]。而且现在通用的无线局域网标准 802.11 标准传输中，最高的 802.11n 标准的理论最高速率可达 600 Mbps[3]。同时 WiFi 因其基站铺设简便、廉价得到广泛发展[4]，被认为是解决移动数据网络拥塞的最主要的候选方案。WiFi 聚合实现从 LTE 网络转移移动流量到 WiFi 网络功能，是一个非常有前景的技术。所以，由于 WiFi 部署的低成本和提供的高速率，使其不仅获得个人和企业的广泛使用，也越来越引起了运营商无法承受的蜂窝网络负载的关注。 许多运营商将运营商级 WiFi 网络作为蜂窝网络的热点，以缓解快速增长的蜂窝网络流量的压力，并改善用户体验。

### 1.1.1 LTE-WiFi 聚合

实际上，十多年来第三代合作伙伴计划(3rd Generation Partnership Project，3GPP)一直致力于蜂窝网络(Cellular Network)与无线局域网(Wireless Local Area Networks，WLAN)的互相配合，从 Release 6 的 Interworking WLAN(I-WLAN)概念到最新的 Release 13 和 14 的 LTE-WiFi 聚合(LTE-WiFi Aggregation，LWA)[5][6]。早期有关互相配合的研究多关注于 WiFi 分流，具体则是完整的鉴权和计费机制（例如使用 SIM 证书保证 WLAN 安全[7]）以及无缝切换(允许用户设备（UE）使用相同的 IP 地址在移动和无线网络间切换)。从 Release 8 开始，3GPP 发展了 ANDSF (the Access Network Discovery and Selection Function)框架，它提供了丰富的支持来实现运营商政策包括提供订阅者特定服务、网络选择和流量路由。3GPP 也解决不同种类的 WLAN 的支持部署(例如，信任

和非可信无线局域网) [8]。除了 3GPP，几个其他标准组织如 GSMA、IEEE WBA、和 WFA 都在跟进无线网络聚合的研究。

到目前为止有关网络聚合讨论的关键限制在于不能在 LTE 和 WLAN 接入之前进行 IP 流聚合的操作。这样聚合系统就可以显著增加用户可以享受的峰值吞吐量并提高系统容量和网络利用率。不像之前的 WiFi 分流中属于一个数据连接的包只可以通过 LTE 或 WiFi 发送，LWA 允许一种数据连接的数据包被分割并通过两种方式发送[4]。在以前对载波聚合和双连接的研究基础上，3GPP 已经探索了类似的体系结构，主要是针对 LTE 中的介质访问控制(MAC)和分组数据融合协议(Packet Data Convergence Protocol，PDCP)。并最终决定在 PDCP 实现这一聚合操作。

LTE-WiFi 聚合系统从架构上看作是 LTE 双连接的系统结构[9]。这种场景下，WLAN AP(Access Point)可以与 LTE 演进型基站(eNB) 集成在一起。LTE eNB 在收到核心网传输下来的数据包后，会考虑诸多对数据传送有影响的因素，比如当前 LTE 与 UE 以及 UE 与 WLAN AP 间的链路状况、eNB 以及与之相连的 WLAN AP 的负载情况等[10]，加以判决，从而动态调整数据包或业务是交给 LTE 网络传输给 UE，还是由 WLAN 网络传输给 UE。如果 LTE eNB 决定将某些数据由 WLAN 网络传输给 UE，那么这些数据包可以通过 PDCP 层进行动态分流，并通过新添加的适配层或适配功能将 LTE 数据包进行适配和包装后发送给与之相连的 WLAN AP[11]，进而由 WLAN AP 发送给相应的 UE，而该 UE 收到 WLAN 网络分流的数据包后，会在 PDCP 侧与经由 LTE 网络收到的数据包进行合并，然后再传给高层。而且，LTE eNB 可以根据反馈的信息对实时的网络状况以及用户的业务状况做出分析，之后对使用的 LTE 与 WLAN 网络资源进行快速、动态的调整，从而达到最大化的利用网络资源的目的 [12]。

## 1.1.2 SDR 技术

为了进一步评估实际的 LTE-WiFi 聚合系统的性能，我们使用软件定义无线电技术作为系统实现平台。这是一种无线电广播通信技术，不同于传统的通过硬连线实现，它基于软件定义的无线通信协议实现。系统所需要的协议、频带和功能模块都可通过软件的下载和更新来升级，而不用完全更换硬件，也就是通过尽可能的在软件中实现通信模块以代替传统通信系统模型中的硬件组成，从而便捷、快速的搭建完整的通信系统[13]。

由于技术本身的优点，大量 SDR 项目正在快速发展，而且这些项目多数都免费提供硬件原理图和软件源代码，这样，开源方式又极大的促进了 SDR 项目的发展，提高了项目的质量。因此，不管对于通信领域的理论研究或者验证实现，开源 SDR 都提供了灵活和高效的手段[14]。在开发 SDR 应用程序时，程序的每个组件都可以单独编写，反复调试，同时结合相应的硬件收发机，便可以模拟真实的通信系统，所以在开源

SDR 平台上设计并实现 LTE-WiFi 聚合系统非常方便[15]。

## 1.2 国内外研究现状

目前大多数智能手机都内置蜂窝和 WiFi 调制解调器，智能手机和移动通信量的增长意味着 WiFi 订阅和流量的巨大峰值。用户对高质量视频内容的强烈渴望使得即使是 WiFi 服务提供商也与移动网络运营商达成了相同的目标：更快的速度和更好的服务质量[16]。面对可用频谱的稀缺性和高额许可费的负担，移动运营商一直在寻找出路，例如通过使用小小区来提高容量。然而，这导致干扰和移动性控制问题，这是昂贵的。所以，移动运营商迫切需要更经济的解决方案来保证更多的频谱可用。

为了应对日益增加的 WiFi 流量，已经安装了更多的 WiFi 接入点，这使得 WiFi 在城市更容易接入。尽管今天的 WiFi AP 支持 Gbps 级的高容量，但是由于多用户共享信道的 WiFi 的性质，仍然很难保证视频流的质量。此外，由于覆盖范围较小，即使在低速移动时也可能出现质量下降或服务中断的情况。这就是为什么 WiFi 服务提供商需要一个解决方案来保证甚至在交通堵塞或移动时也能保证最低的服务质量。

LTE-WiFi 聚合技术作为分担运营商数据流量压力的重要补充和提升网络服务质量的重要途径而受到广泛，目前的研究主要从以下几个方面进行总结：

标准方面： 3GPP 在 Release 14 中提出了 LTE-WLAN Aggregation，标准已经对下行 WiFi 分流的做出了明确的定义，从整体的无线协议结构到网络接口，以及 WLAN 的测量、认证都有规范的说明。通信行业的研究者可以清楚的遵循协议标准进行开发实现。

实现方面：目前，关于 WiFi 分流的理论比较完整，包括不同的聚合层次，分流实现以及各自的优缺点。尤其是不涉及 LTE PDCP 层[17]、RLC 层(Radio Link Control，无线链路层控制协议)[18]的高层聚合由于实现比较简单，一些知名的研究机构和企业已经做出了基于相关技术的试验产品[19]。比如韩国的三星电子在移动端基于 MultiPathTCP(MPTCP)[20]实现了移动网络流量和 WiFi 流量的融合，使得客户端速率得到极大提升。而作为 LTE-WiFi 聚合的重要推动者，高通也发布了 LTE-WiFi 聚合系统实现的验证视频。

OpenAirInterface(OAI)系统是欧洲EURECOM组织为实现开源LTE生态系统而提供一个灵活稳定的研究平台[21]。其中 OAI 中已经实现了很多关于 LTE 相关的链路级仿真实验，也能基于硬件设备实现各种通信协议。通用软件无线电平台(Universial Software Radio Peripheral，USRP) 是针对世界范围无线电通信领域多体系并存，不同体系间通信设备不兼容 ,无法订制统一标准等实际问题的情况下产生的。作为一种潜在的实际解决方案，它通过软件来建立一种灵活的无线电系统，可实现多种服务，适应多种标准，覆盖多个频段，而且软件编程，代码的重构和修改更加便捷灵活，使用户也可以通过下载更新 ，随时随地适应他所在的无线电环境。刚好可以使用多个 USRP 用来搭

建一个实际的一发一收的模拟测试环境来研究 LWA 技术的性能，与实际的仿真结果进行一个比较，看实际的测试效果是否与仿真结果是一致，同时使用 USRP 来进行搭建环境也存在很多的优势，搭建测试环境的时间短，进行试验的成本更低。目前基于 USRP 和 OAI 系统已经完成了很多的新技术协议的研究，如 Agostini 实现了一种智能交通系统专用的短程通信技术（DSRC）协议[22]，Ngoc-Duy Nguyen 实现了一种用于 LTE 的多媒体广播和多播的技术[23]，Bassem Zayen 实现了空间交织认知无线电技术[24]，同时 OAI 还能为 5G 研究提供一个良好的研究平台[25]，方便验证通信新技术[26]。

## 1.3 论文内容安排

本论文共分为五章，文章结构和各章节的主要内容安排如下：

第一章，为本文的绪论部分。首先介绍了本文的研究背景；然后对本文中研究的 LTE-WiFi 聚合技术和 SDR 技术的发展背景与应用现状做了简单介绍；最后，说明了本文的文章结构及主要内容。

第二章，主要对 WiFi 融合技术的发展以及 SDR 技术的应用做了介绍。首先，简要介绍了 WiFi 融合技术的理论发展，通过几种不同技术的对比对 LWA 技术进行了分析。然后，简单介绍了 SDR 技术以及基于 SDR 的几种软件平台，并着重讲述了 OAI 平台的优势。最后，对于本文用到的硬件设备进行了说明。

第三章，介绍了基于 SDR 的 LTE-WiFi 聚合系统的设计与实现。首先详细介绍了 LWA 技术原理，然后分别介绍了基站侧和用户侧的工作流程和主要模块功能的实现，最后搭建实验环境，得出了实验结果并进行整理分析。

第四章，提出了一种 LTE-WiFi 聚合系统的分流策略优化实现。通过在基站侧修改分流策略实现了流量的动态调整。然后通过实验对分流策略进行了性能分析。

第五章，总结与展望。首先对本文提出的 LTE-WiFi 聚合系统的功能进行总结。然后说明了该工作目前仍然存在的问题以及对下一步工作的展望。

# 第二章　WiFi 融合系统概述及软件无线电技术应用

## 2.1　WiFi 融合系统概述

由于全球移动数据业务量的爆炸式增长，移动运营商必须应对前所未有的困境，在增加移动数据业务的同时需要保持低成本。作为一个可能的解决方案，越来越多的运营商开始研究使用 WiFi 分流来减少网络基础设施的负载，同时保证为用户提供的服务质量。除了增加网络容量，WiFi 分流还可以提供更好的覆盖范围，尤其是当用户在小区的边缘附近或者在网络覆盖较弱的室内环境中。WiFi 网络的高可用性和相对低成本也引起了移动运营商对 WiFi 分流的热切关注。

3GPP 组织一直致力于蜂窝网络与无线网络的互相配合，从最早在版本 6 中提出的 I-WLAN 概念（主要内容是通过电路交换 UTRAN（UMTS Terrestrial Radio Access Network）接入传输用于语音服务的 GSM 信令，而通过 WiFi 接入传输用于语音业务的 SIP 信令），到最新的版本 13 和版本 14 中提出的 LWA。早期有关使蜂窝网络与无线网络互相配合的努力主要是在 WiFi 分流上，专注于集成认证、计费，（例如使用 SIM / USIM 证书保证 WLAN 安全）以及无缝切换（如使用演进的分组数据网关节点时如何允许 UE 可以在 Cellular 和 WiFi 之间切换时使用相同的 IP）等主题。从版本 8 开始，3GPP 开发了接入网络发现和选择功能(ANDSF)框架，它的出现为实现运营商政策提供了有力的支持，比如提供订阅者特定服务，网络选择和流量路由功能。3GPP 也解决对于不同种类的 WLAN 的部署的统一支持，包括可信任无线局域网和非可信无线局域网。除了 3GPP 之外，其他的几个标准组织如 GSMA、IEEE WBA、和 WFA 也参与了这一工作，做出了许多贡献。最近一段时间，3GPP 的研究方向则是通过在无线接入网络(RAN)层面利用可用的实时信道和业务信息来提高 WiFi 分流的性能，同时减少 WiFi 融合对于蜂窝核心工作网络的影响。之后，为了支持基于无线电和负载测量的 LTE/UMTS 和 WiFi 之间的流量转向，3GPP 在版本 12 中发布了以 UE 为中心的 RAN 辅助 WLAN 互通功能，在版本 13 中发布了以网络为中心的 RAN 控制 LTE-WLAN 互通功能[27]。

近来，已经引入了无线链路层面和 TCP 层面的新型 LTE-WiFi 聚合解决方案[27]。第一类无线链路级聚合包括 LAA( License Assisted Access)和 LWA，目前 3GPP 正在对其进行标准化。 LAA 和 LWA 的优势在于它们可以同时使用 LTE 和 WiFi 频带进行数据传输，并且由于 LTE-WiFi 聚合发生在设备和基站上，所以不需要单独的 WiFi 接入 GW(网关，GateWay)。 第二种是 TCP 级聚合，其中包括目前 IETF 正在讨论的基于多路径 TCP（MPTCP）基于代理的聚合。 MPTCP 可以拆分数据并同时通过 LTE 和 WiFi

网络传输。 LTE-WiFi 聚合发生在设备和 MPTCP 代理上，因此不需要专门的网络设备进行聚合。

对于 LTE 业务传输，LWA 使用与 LAA 一样的未授权频段，但是与 LAA 不同，通过 WiFi 进行传输。这意味着 LWA 不需要新的支持 LTE 的 5 GHz 硬件，并且可以通过连接到 LWA 基站的 WiFi AP 传输 LTE 业务。此时，WiFi AP 可以在没有专用 GW 的情况下使用 LTE 核心网络功能（例如认证，安全性等），所有这些都不会干扰本地 WiFi AP 甚至是轻微的[28]。

LWA 体系结构由 LWA eNB，LWA WiFi AP 和 LWA UE 组成。 LWA eNB 和 WiFi AP 可以并置或不并置。与 LAA 不同，LWA 在 LTE 频段上使用 LTE，在 WiFi 频段上使用 WiFi。这消除了 WiFi 频带上潜在的公平性或规定问题。然而，由于 LTE 数据必须在 eNB 处分裂然后在 UE 处聚合，所有涉及诸如 eNB，WiFi AP 和 UE 的节点都必须是 LWA 使能的。当然，LWA 体系结构，协议和操作也必须被定义。下面我们将比较 LWA 的潜在问题和优势。

成本：与 LAA 不同的是，新硬件不是必需的，现有的基站和 WiFi AP 可以通过软件升级来实现 LWA，这比 LAA 需要安装新的小型基站更为经济，特别是在大规模部署的情况下。但是，这种解决方案的缺点是需要能力较差的 eNB 或者小型小区所需的投资成本，并且 WiFi AP 需要被替换，或者在本地 WiFi AP（运营商部署，用户部署，市政 Wi-Fi 等）的情况下）不能使用。成本可能不如 LAA 那样高，但仍然可能是显着的。

设备：应该向 UE 添加 PDCP 重新排序和 PDCP 聚合特征。 UE 可以利用当前可用的 LTE 和 WiFi 调制解调器。因此，传统的 UE 也可以被认为是对于 LWA 兼容的，但只有在它们的软件可升级的情况下。

标准化状态：对于 WiFi 融合的标准化工作一直在进行，直到版本 14 中，3GPP 提出了 LTE-WiFi 聚合系统的概念，包括部署、测量和安全等各个方面的特性。在这之前 WiFi 接入讨论的关键限制在于所有的方法都不允许在 LTE 和 WLAN 的 IP 流聚合，这样的结果是属于一个数据连接的包只可以通过蜂窝网或者无线网络发送，但是在 LWA 中，系统允许一种数据连接的数据包被分割并通过两种方式发送。在以前研究载波聚合和双连接技术时，3GPP 已经探索了类似的集成结构，主要是针对 MAC 和 PDCP 层[29]。在 LTE 和 WLAN 互通的情况下，这样的集成是具有挑战性的，因为这些技术是有明显区别的，并且存在大量已经部署的 WLAN AP 基站，需要无缝地与任何所提出的互通解决方案一起工作。

具体的，在 LWA eNB，DL 用户业务在 PDCP 层被分割，然后通过 LTE 和 WiFi 转发。一些 PDCP 分组通过 LTE 无线电链路上的数据无线电承载（DRB）传送，而另一些 PDCP 分组被发送到 WiFi AP，由 eNB 添加 DRB ID 以指示它们属于哪个 DRB 并传送 LWA PDU 通过在 Xw 上建立的 IP 隧道连接到 WiFi AP。 WiFi AP 然后将 Ethertype

设置为 PDCP，并通过 802.11 接口将 LWA PDU 转发给 LWA UE。在接收到 802.11 帧时，如果 Ethertype 设置为 PDCP，则 LWA UE 将帧转发到 LTE PDCP 层。然后，PDCP 层通过检查其 DRB ID 来收集从 LTE 和 WiFi 接收的属于相同 LWA 承载的 PDCP 分组，并且通过重新排序来聚合它们。 LWA 适配协议（LWAAP）定义为 3GPP TS 36.360，以支持这些 LWA 操作

综上所述，在无线链路层执行 LTE-WiFi 聚合的 LWA 不需要为核心网增加额外的功能，但需要安装新的或者升级的 RAN 设备和设备。不会导致公平访问问题，因为它只允许 WiFi 使用未经许可的频段。必须修改协议以在 PDCP 层处聚合 LTE 流量。LWA 的成本效益将取决于设备和设备是否可升级。而基于 SDR 平台实现的 LWA 则可以很好的解决系统的可扩展性和灵活性问题，我们将在下一小节详细介绍。

## 2.2    SDR 技术与软件平台 OAI 概述

软件定义无线电是一种无线电广播通信技术，它基于软件定义的无线通信协议而非通过硬连线实现。SDR 概念由从 Joseph Mitola 博士于 1992 年提出，它的核心思想是通过建立一套高度模块化的通用硬件平台，将通信系统中的大部分功能包括频带处理、空中接口协议和调制编码等功能通过软件模块实现，这样，之后移动通信系统的更新升级只需要在软件模块中更新代码就可以完成，具有很强的开放性、可扩展性。

由于软件编程相对于硬件编程在可移植性、调试难易度及可扩展性上的天然优势，伴随着通用计算机平台性能的不断提升，SDR 技术也获得了越来越多的重视，尤其是一些国际机构和组织对 SDR 平台开发与应用的研究，取得了不错的成果。目前最常用的 SDR 系统通常由通用计算机设备和连接在计算机上的外部设备组成。著名的外部设备有 Ettus 公司的通用软件无线电外设、Nuand 出产的 BladeRF，以及 Great Scott Gadgets 生产的 HackRF，此外，还有性能较差但是价格低廉的 RTLSDR。本文选用了 USRP 作为 LWA 系统基站和用户侧外接的硬件设备，具体内容将在后续的小节中详细介绍。

外部设备迅速发展的同时，基于 SDR 的软件平台也逐步建立起来，其中就有许多实现了 LTE 系统的开源软件平台，这里边比较著名的有 OAI，OpenLTE 和 SRS(Software Radio System)。在文献[16]上，作者成功运行了 OAI 和 OpenLTE 系统，并且从 LTE 功能和运行的稳定性等方面对比了两套程序。OAI 程序基于 LTE 协议 Release 8 和 Release 10 都有完整的系统。它能够运行频分双工（Frequency Division Duplex，FDD）和时分双工（Time Division Duplex，TDD）两种模式，有完整的上下行物理信道，支持 TM1(Transmission mode，传输模式)、TM2、TM4、TM5 和 TM6，有信道质量指示（Channel Quality Indicator，CQI）和预编码矩阵指示（Precoding Matrix Indicator，PMI）上报功能，基本上实现了比较完整的一套长期演进系统，而且可以基

于是否带核心网有两种编译运行方式，而 OpenLTE 则只是实现了 LTE 协议的部分功能。SRS 系统则是在 OpenLTE 的基础上添加完善部分功能，实现了 TM1 和 TM2 的功能，用户侧更加稳定，但相比较 OAI 系统，整体上仍有不足。综合对比后，本文选择了 OAI 平台作为实现 LWA 系统的软件平台[31]。

OpenAirInterface，又称 OpenAirInterface5g，是欧洲 EURECOM 组织发起并维护的一个开源 SDR LTE 项目。同时也是目前最全面，最具竞争力的开发平台之一，它根据 3GPP 的标准，完全实现了 LTE 协议的核心网（EPC），基站（eNB）和用户（UE）三部分，目前已经支持 Release 10 的功能，并且在持续更新，其软件目录如图 2-1 所示。

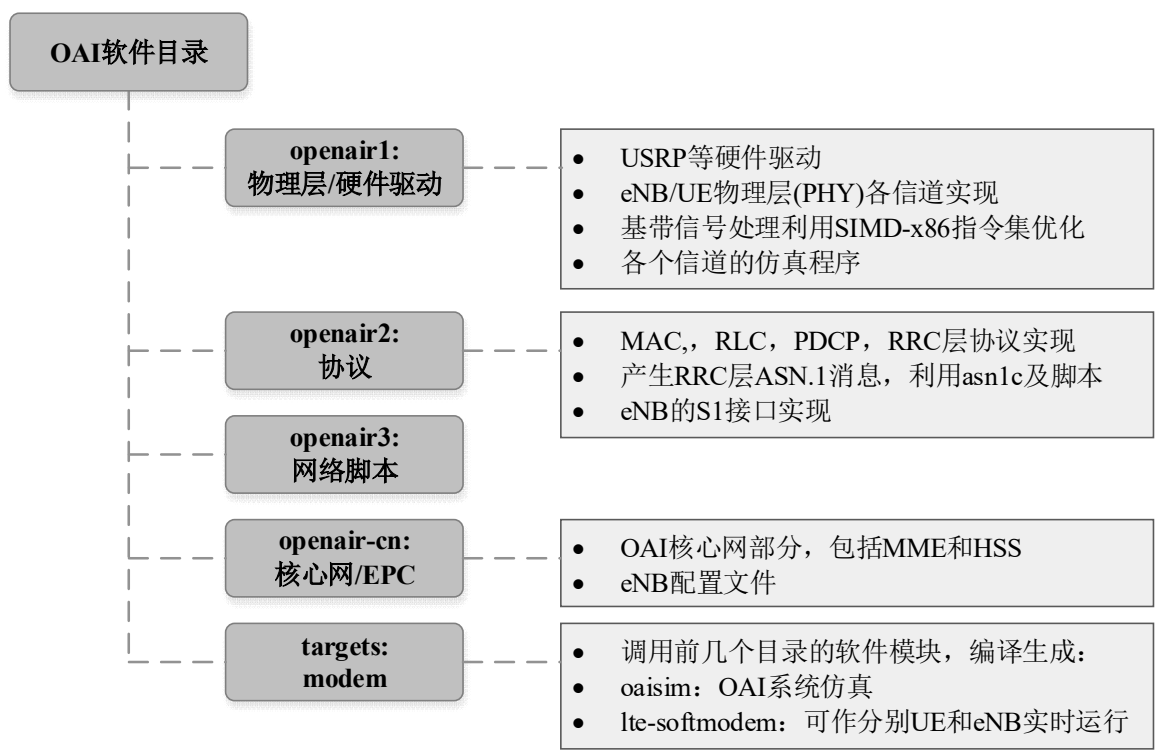

图 2-1 OAI 的软件架构

## 2.3 系统硬件设备概述

USRP 是由 Ettus 公司研发的一种低成本的 SDR 平台。它功能强大，开发者可以通过编写 UHD（USRP Hardware Driver，USRP 硬件驱动）实现多种框架，获得各种各样的应用。本文选择了 USRP B210 作为通用无线电外设[30]。它的结构如图 2-2 所示。

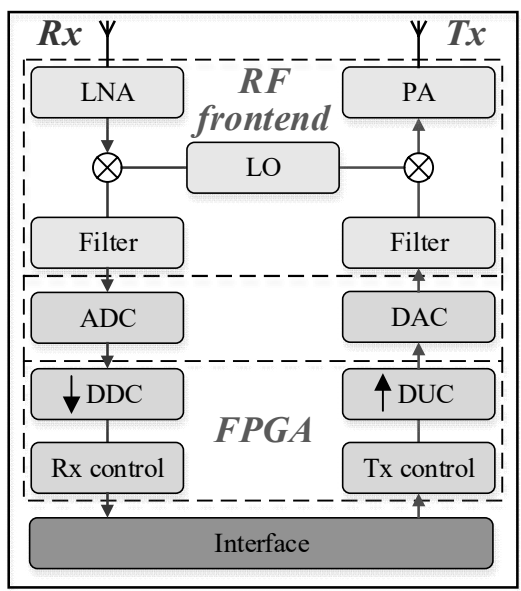

图 2-2 USRP B210 组成图

在接收侧，通过 USRP 的射频前端收到的信号进过一系列处理后由 USB3.0 接口传入 GPP，在 GPP 进行基带信号处理。在发射侧，与之相反，通过 GPP 处理产生的数字基带信号经过 FPGA 和 DAC 后变成模拟信号，然后经过滤波器和振荡器变为射频信号，最终经过功率放大器（Power Amplifier，PA）由天线发射出去。

## 2.4 本章小结

本章首先简要介绍了 WiFi 融合技术的发展过程，然后介绍了 SDR 技术和系统使用的软硬件平台，然后详细介绍软件平台 OAI 和硬件平台 USRP B210，分别对其的功能和框架进行了总结，这是实现本文工作的基础。下一章将介绍如何在 OAI 平台上实现 LWA 系统。

# 第三章 基于 SDR 的 LTE-WiFi 聚合系统设计与实现

## 3.1 LTE-WiFi 聚合原理概述

长期以来，对于 WLAN 与 LTE 融合相关技术的研究主要都是集中在核心网一侧，但是对于运营商来说，基于核心网的融合机制并不能充分实现对网络的灵活控制，例如接入网络的动态信息（例如网络负载、链路质量、回传链路负荷等）对于核心网都难以实施监督。为了使运营商能够得到 WLAN 和 LTE 网络使用情况的更加及时的反馈，从而加以判断，采取更加灵活、更加动态的联合控制，达到进一步降低运营成本、提供更好的用户体验的目的，同时也能更有效的利用现有网络、并降低由于 WLAN 持续扫描造成的终端电量的大量消耗。因此，3GPP 组织近年来对无线网络侧的 LTE/WLAN 互操作方式越发的重视起来，展开了相关的研究以及标准化工作。

3GPP LTE 项目研究了 LTE 和 WLAN 互通，以解决版本 12 中数据流的链路选择问题。 但是，数据流仍然只能选择一方，3GPP 版本 13 研究了 LTE 和 WiFi 数据聚合，即将基站的 PDCP 中的分包数据处理，并分别从 LTE 和 WLAN 接入点（AP）传输到 UE。 最近，3GPP 标准发布了完整的 LWA 系统架构。

如前所述，通过参考 LTE 双连接的系统架构，就可以更容易地理解 LWA 的系统架构。在 LTE 与 WLAN 共址融合场景下，WLAN AP 可以与 LTE 基站集成在一起，而在 LTE 与 WLAN 非共址融合场景下，WLAN AP 可以通过 Xw 接口与 LTE eNB 相连接。LTE 基站侧在收到上层发送过来的数据包后，会考虑根据实际状况，比如当前 LTE 基站的负载情况、基站相连的无线路由器的负载情况以及 LTE 和 WiFi 传输的链路状况等，来判断这些业务数据是由 LTE 网络传输给 UE，还是由 WLAN 网络传输给 UE。对于 LTE 基站决定由 WLAN 网络传输给 UE 的数据包，可以通过基站侧 PDCP 层进行动态分流，并通过新添加的适配层或适配功能将 LTE 数据包进行适配和包装后发送给与之相连的无线网络，进而通过 WiFi 发送给相应的 UE，而该 UE 收到无线网络分流的数据包后，会与经由 LTE 网络传输收到的数据包在 LTE 用户侧的 PDCP 层进行合并操作，处理后再传给高层。而且，LTE eNB 可以根据实时的网络状况以及用户的业务状况对使用的 LTE 与 WLAN 网络资源进行快速、动态的调整，以达到资源利用率的最大化。

此外，3GPP 还提出过另一个非授权频段的技术，授权辅助接入技术（LAA），但与之不同的是，在 LWA 中，LTE 依旧运行在授权频段上，LTE 无需针对非授权频段的特性及法规要求进行重新设计，更方便的是，从实际操作的角度讲，LWA 只需通过简

单的软件升级启用[32]，而不需要针对基站和终端进行硬件的改造升级，这样避免了大规模的改动[33]。同时，WLAN 网络也可以借助自己十分成熟的针对非授权频段的接入和共享机制，依旧工作在非授权频段上。而且，随着 WLAN 技术的飞速发展，包括超高速超高频段 WLAN 系统的成熟与应用，比如工作在 60 GHz 的 802.11ad 系统，这两种无线技术通过 LWA 的结合将会为用户带来绝佳的用户体验。

LWA 解决方案支持两种数据承载：分流承载和切换承载。用户平面的协议体系结构如图 3-1 所示，顾名思义，属于 LWA 切换承载的报文总是由 eNB 调度通过 WiFi 进行发送，而属于 LWA 分流承载的报文可以通过 LTE 或 WiFi 进行调度。 对于分流承载，数据流的传递过程如图 3-1 所示，在 eNB 数据到达 LTE PDCP 之后，PDCP 将部分数据分组布置到 RLC 并由 LTE 网络发出。 另一部分则交付给 LWA 适配协议层，并根据安排由 WiFi 网络发送。 在 UE 侧，PDCP 把 LTE 和 WiFi 接收后的数据包一同进行处理，不过，在向 PDCP 发送数据之前必须进行排序处理，这是由于 LTE 与 WiFi 之间的网络传输时差导致数据包不能按顺序到达用户侧。 最后，将 PDCP 中处理的数据发送到较高层，完成整个网络链路的数据传输。

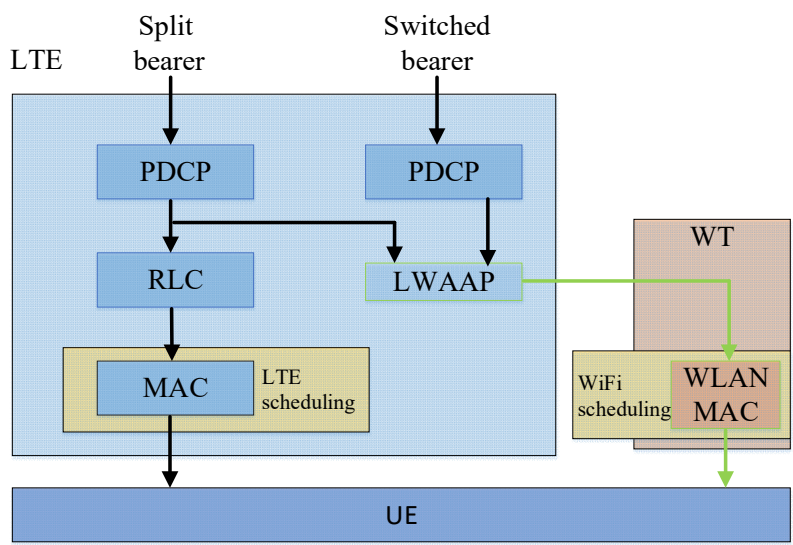

图 3-1 LWA 系统结构图

实际上，现在主要有以下三种 LTE-WiFi 融合架构：

（1）松耦合模型，WiFi 和 LTE 网络在物理上是独立的但是他们都连接到相同的 IP 网。为了使用 WiFi 网络，用户设备首先需要扫描所有可用的无线接入节点，然后需要在选择的接入点上进行验证鉴权，之后才开始发送或者接收数据。这样，用户设备在需要分流流量时就不得不花费时间在无线接入点。

（2）紧耦合模型，无线接入节点直接连接到 EPC。然而，用户设备仍然需要使用 WiFi 侧的安全机制，也因此需要付出相应的时间。

（3）非常紧耦合模型，也就是 3GPP R13 发布的 LTE-WiFi 聚合模型。被 LTE eNB

覆盖的无线接入节点直接连接到 eNB。数据业务就可以直接被分流给 WiFi 同时其它安全，移动性等控制功能交给 LTE 网络保证。

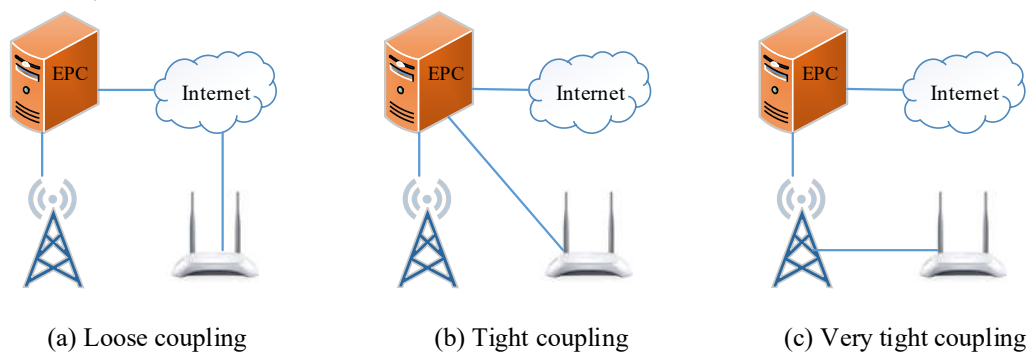

(a) Loose coupling      (b) Tight coupling      (c) Very tight coupling

图 3-2 三种 LTE-WiFi 聚合方式

如图 3-2 所示，LTE 与 WiFi 之间的耦合非常紧密[10]，与 LWA 类似，其他都是松散的联轴器和紧耦合。在松散耦合模型中，WiFi 和 LTE 在物理上是独立的，并且它们连接到相同的 IP 网络。为了使用 WiFi 网络，UE 需要扫描所有可用的无线接入节点，然后在发送或接收数据之前需要验证选择和接入点认证。以这种方式，当 UE 需要分流时，UE 必须花时间在 WLAN AP 中注册、切换。在紧耦合模型中，无线接入节点直接连接到演进分组核心。然而，UE 仍然需要在 WiFi 侧使用安全机制，当然还需要支付相应的时间。在非常严格的模型中，无线接入节点直接连接到 LTE 覆盖的 eNB。数据业务可以直接转移到 WiFi。同时，其他安全机制和移动控制功能也得到了 LTE 的保障。所有 WiFi 流量都可以从移动运营商 EPC 提供的服务中受益。这些服务包括计费，深度包检测，合法拦截，策略，认证等等，这些列表继续进行。如果 LTE 信号丢失，则该链路将被丢弃，用户可以通过 WiFi 连接到互联网。一般来说，紧密耦合更适合于三种技术中运营商的需求。

在本文中，基于 OAI 实现了 LWA 系统。 考虑到早期 OAI 实现中的不稳定性和比特率非常有限，进行了一些修改以提高系统性能。 特别是通过添加虚拟网络适配器和共享内存来优化从本地网卡（NIC）到 LTE 无线链路控制层的过程。 并且重排序功能被设计为将有序数据分组转换为 UE 侧的 PDCP，这对于应用的流量是非常重要的。 此外，有序数据避免了早期研究中 PDCP 重排序定时器对上层的不良影响。 因此，建立了包括 LTE 数据链路和 WiFi 数据链路在内的整个数据链路。 在实验中，系统运行良好，具有高速稳定性，带宽为 5 MHz 的 LTE 的测试速率达到 10 Mbps 以上。 此外，WiFi 显示在演示中清除 LTE 中流量的重要作用。

## 3.2 开发环境搭建

### 3.2.1 硬件平台

我们的测试台如图 3-7 所示。在实验场景中部署了两台具有 SDR [11]外设的计算机，分别被认为是 eNB 和 UE。 同时，eNB 连接无线路由器，作为 WLAN AP。 无线网络适配器插在 UE 的计算机上。 在 eNB 和 UE 之间部署 LTE 和 WiFi 完成。 SDR 是开放，可扩展性和兼容性的一个通信平台。 我们可以加载不同的软件，在这个通用硬件平台上实现不同的通信功能。 通过使用 SDR，我们可以快速更改频道接入或调制。通过使用不同的软件可以适应不同的标准，构成高度灵活的多模手机或多功能基站。

### 3.2.2 软件平台

根据 OAI 官方网站的建议，对本文设计所需的开发环境进行了配置。OAI 软件需要运行在 Intel 架构的中央处理器（Central Processing Unit，CPU）上，并且需要能够支持单指令多数据流 (Single Instruction Multiple Data, SIMD)指令集，我们选择了 intel i7 CPU 来实现。 由于 OAI 程序所需系统资源巨大，GPP 的物理内存需要至少 8G，同时 GPP 需要有 USB3.0 接口与 USRP B210 进行通信，如果不慎使用了 USB2.0 接口通信，则会出现数据处理过慢的问题，导致程序崩溃。为了实现实时通信，还需要对 GPP 系统进行配置。首先，需要安装标准版 64 位的 Ubuntu 14.04 LTS 系统，采用 Linux 系统内核 3.19 版本。然后，按照以下步骤编译内核相关内容：

● 在 BIOS 设置中关闭所有功率管理选项以及 CPU 的变频功能；

● 输入以下代码，更改内核内容，此过程需要 GPP 连接入互联网。

```
version=3.17
wget    -r    -e    robots=off    --accept    -regex"(.*lowlatency.*amd64)|(all).deb"
http://kernel.ubuntu.com/~/kernel-ppa/mainline/v${version}-utopic/
dpkg -i kernel.ubuntu.com/*/*/*/*deb
ln        -s        /usr/src/linux-headers-${version}*lowlatency/include/generated/autoconf.h
/lib/modules/${version}*lowlaltency/build/include/linux
```

● 重启 GPP，输入 uname -a 命令，查看自己电脑内核是否到 3.17-lowlatency 版本。如果失败，需要按照上述步骤重新操作。

接着需要安装 UHD 软件，用来驱动 USRP。然后通过命令

```
git clone https://gitlab.eurecom.fr/oai/openairinterface5g.git
```

来安装 OAI，最后通过命令./build_oai –I，自动更新 OAI 所需依赖库，完成 OAI 平台所要求的开发环境配置。

此外，由于在带有 L2 进行测试时，OAI 平台表现出很差的稳定性，所以本文参考相关协议规定，根据系统的要求，对 L2 部分代码尤其是数据面相关的 RLC 层功能做了修改。具体如下一小节所述。

### 3.2.3  OAI 协议优化

下面将结合实现中的附图，对这部分协议栈修改的技术方案进行详细地描述。PDCP 层处理之后，在 RLC 层分段或拼接后的数据需要经过 MAC 层的调度，进而到达 PHY 层，为了解决原有的 OAI 代码中对这一复杂协议结构实现导致的系统不稳定，本文参考相关协议和原有代码，主要针对 RLC 层部分，提供了另外一种的实现结构[36]，这里首先对本文重构的数据传输结构进行详细说明。

（1）发送端

对实现的协议结构进行详细说明，具体流程如下：

**步骤 1**，从 PDCP 获取上层数据，其中，上层数据是长度不同的多个数据。首先需要说明的是，这里的发送端和接收端，指的是也就是 OAI 的基站侧和用户侧，发送数据的发送方和接收数据的接收方，实际应用过程中，发送方可以接收终端发送的请求。当然在上行链路中，发送端可以是用户，接收端可以是基站，但是这里我们只研究下行过程。

基站获取到用户的业务请求，将该业务请求进行处理，得到业务请求对应的上层数据。面向业务，一般都是在应用层进行的，所以实际的过程中，基站是在应用层获取到用户的业务请求，并对该业务请求进行处理的。在数据通信的过程中，得到的上层数据是长度不同的多个数据，不是每一个数据都正好与底层要求匹配，从而使底层可以传输该数据，所以传输该上层数据首先需要对该上层数据进行处理，使得上层数据与底层要求长度匹配。

**步骤 2**，分段或拼接上层数据，得到对应与底层要求长度相匹配的待发送数据。在 RLC 层的功能中，对于上层数据，当物理层提供的物理资源相对该上层数据的长度较短时，需要对上层数据进行分段。由于不是每个分段都能正好匹配物理层提供的传输比特长度，因此需要对某些分段添加填充比特。对于变长数据的传输，为了可以在接收端进行处理，一般除了传输数据内容外，还需要传输数据包头，数据包头中包含对数据内容的控制信息，如数据分段长度、填充比特长度等等。

数据通信中，当每次传输的上层数据比较长时，都包含一个数据包头和一个数据内容分段，接收方根据数据包头指示接收处理。当传输的应用程序比较短时，如果每次传输都只包括一个数据包头和一个数据内容，则传输效率会很低。因此可采用一个数据包头指示多个数据负载分段的方法，减少由于数据包头控制信息对物理资源的占用，提高数据传输的效率，所以协议结构实现在发送端对上层数据进行分段或拼接操作，生成与底层的资源要求长度相匹配的待发送数据，使得经过空口传递后，接收端根据数据包头的控制信息进行重组，恢复原有的上层数据。通过对数据进行分割和拼接操作，能够充分利用底层资源，提升传输数据效率。

**步骤 3**，将待发送数据转发至发送端的物理层，并从发送端的物理层，将待发送数据发送至接收端的物理层。将上层数据处理完成后得到的待发送数据是与底层相匹配的数据，直接将待发送数据从发送端的物理层发送至接收端的物理层。为了使数据传输的过程更加方便以及安全，将待发送数据从发送端的物理层发送至接收端的物理层之前，在协议结构的一种实现方式中，可以在发送端的物理层将上层数据进行调制、编码以及加扰等处理。

本文实现的协议结构，将上层数据进行分段或拼接的处理，得到与底层匹配的待发送数据，分段拼接后的待发送数据经过 MAC 层发送至接收端，然后直接通过发送端的物理层发送至接收端，简化了原来数据传输的结构代码。

在这种协议结构的实现中，分段或拼接上层数据，得到对应与底层相匹配的待发送数据，包括：

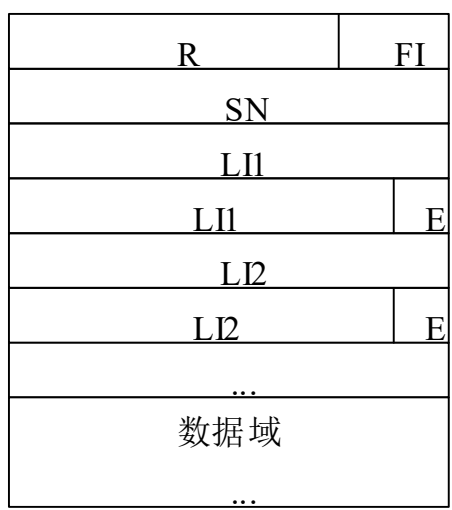

图 3-3 自定义协议头

根据预设协议，将上层数据分段或拼接为包括数据包头和数据内容的待发送数据，其中，数据包头包括：预留比特位、第一标识位、待发送数据序列号位、第一长度位、第一扩展位、第二长度位以及第二扩展位。

其中，预设协议是自定义的协议，定义各个标识表示的含义，即对数据内容的指示信息。对本文实现定义的数据包头中的各个标识位进行详细说明如下：

预留比特位为扩展位，为本文实现协议结构为分段或拼接数据提供了预留资源，如此使得提供了本文实现协议结构中分段或拼接的扩展性；

第一标识位表示转换后的待发送数据的尾部是否为一个应用数据的尾部，转换后的待发送数据的开头是否为一个应用数据的开头；

待发送数据序列号位表示转换后的待发送数据的序列；第一长度位表示待发送数据中第一分段的数据的长度；

第一扩展位表示待发送数据中第一分段后是否拼接其他应用数据；

第二长度位表示待发送数据中第二分段的数据的长度；

第二扩展位表示待发送数据中第二分段后是否拼接其他应用数据。

数据包头如图3-3所示，接下来对其进行详细说明。一个字段长度为一个字节，第一字段前6个比特R为预留资源；接下来的2个比特FI分别代表生成的待发送数据的数据内容的结尾是一个上层数据的结尾、生成的待发送数据的数据内容的开头是一个上层数据的开头，例如当待发送数据的数据内容的第一字节是上层数据的第一字节，最后一个字节不是上层数据的最后一个字节，则FI标识为01。第二字段是待发送数据的序列号SN；之后每两个字段代表待发送数据中的一个分段，每两个字段中的前15个比特表示分段的长度LI，最后1个比特表示扩展位E，E的值为0时，表示其后紧接的是数据内容，E的值为1，表示其后紧接的是另一个分段。需要说明的是，生成的待发送数据的可以是一个分段组成也可以是多个分段组成，具体有多少个分段组成，是在实际应用过程中，根据上层数据的长度以及信道能够处理的数据的长度进行具体处理以及匹配的。这样通过添加自定义的数据包头，对上层数据进行分段或拼接，充分利用了信道物理资源，而数据包头长度一般只占用几个字节，这样就可以提高系统数据传输的效率，同时，协议结构具有灵活性和可扩展性，该方法有潜力为研究、验证通信新技术提供便利。

将上层数据通过分段或拼接，生成和底层相匹配的待发送数据，然后将待发送数据发送至接收端的物理层。然后，将该待发送数据发送至接收端的物理层，包括：通过发送端软件无线电外设，将该待发送数据发送至接收端的物理层。

为上层数据添加数据包头：在数据包头的第二字节添加待发送数据序列号SN，每个待发送数据递增1，假定最大为127，到最大后从0开始循环。判断生成的待发送数据的第一个字节是否为一个上层数据的第一字节，添加第一标识位。获取一个上层数据，如果上层数据的长度大于待发送数据可用数据域的长度，将上层数据写满待发送数据的数据域，记录数据域长度为LI1，由于没有其它分段，将扩展位E设置为0，判断待发送数据的最后一个字节是否为一个应程序数据的最后一个字节，添加第一标识位。记录该上层数据剩下的长度，并将生成的待发送数据放入队列。

如果该上层数据剩下的长度大于下一个待发送数据可用数据域的长度，重复为这些待发送数据添加数据包头的操作。如果上层数据剩下的长度不足一个待发送数据可用数据域的长度，将上层数据剩下的片段填入待发送数据的数据域，并记录相应的分段长度LI。由于待发送数据还未填满，该段扩展位E设置为1，获取下一个上层数据，如果上层数据的长度正好等于待发送数据剩下的数据域长度，将上层数据填入待发送数据的数据域，并记录相应的LI，将扩展位E设置为0。此时由于待发送数据的第一字节不是一个上层数据的第一字节，最后一个字节是一个上层数据的最后一个字节，该待发送数据的FI字段为10。

将所有的上层数据进行分段或拼接完成后，生成的待发送数据已经存放于共享内存中的队列，从待发送数据的队列中取出一个待发送数据，传送给发送端的物理层，经过该发送端的物理层的调制、编码以及加扰等处理后，通过 USB 接口将调制后的待发送数据传输至通用无线电外设，由通用无线电外设发射至接收端的物理层。

（2）接收端

发送端将待发送数据发送给接收端，接收端需要经过处理，从接收到的待发送数据解析出原来的上层数据，对本文实现协议结构中接收端接收到待发送数据后，对待发送数据的处理进行详细说明，包括：

**步骤 1**，接收端的物理层接收来自发送端的物理层的待发送数据，其中，待发送数据包括数据包头和数据内容，数据包头包括：预留比特位、第一标识位、待发送数据序列号位、第一长度位、第一扩展位、第二长度位以及第二扩展位。

通过用户侧通用软件无线电外设，接收端的物理层接收待发送数据可以理解的是，接收端接收的待发送数据就是经过发送端处理后得到的待发送数据，所以待发送数据与上述的待发送数据是相同的，所以这里就不再赘述。

**步骤 2**，根据待发送数据的数据包头，解析或重组待发送数据为上层数据。

对应于上述在接收端将上层数据经过分段或拼接，生成待发送数据的过程，将该待发送数据解析或重组得到原来的上层数据，是根据生成待发送数据时的自定义的数据包头的协议，解析或重组该上层数据。

如前所述，接收端在这里是用户端。用户端接收到发送端发送的待发送数据，因为在实际的应用过程中，在发送端一般都会对数据进行调制、编码以及加扰等的过程，所以对应的，在接收端就需要进行解调、解扰和译码的处理。用户端将经过解调、解扰和译码后的待发送数据存入队列，等待后续处理。

根据待发送数据的数据包头，解析或重组待发送数据为上层数据，包括：根据第一标识位，判断数据内容的结尾是否是一个上层数据的结尾、数据内容的开头是否是一个上层数据的开头，得到第一判断结果；

根据第一扩展位和第二扩展位，判断数据内容是否有多个分段，得到第二判断结果；

根据第一判断结果和第二判断结果，按照待发送数据序列号位的指示，解析或重组待发送数据为上层数据。

解析待发送数据，即根据数据包头的指示信息，从待发送数据的数据内容中得到上层数据，具体的过程包括：获取待发送数据，读取待发送数据的第二字段，进而获得待发送数据的序列号；然后通过读取待发送数据的扩展位 E，判断待发送数据内的数据域是否存在拼接的情况，记录数据域的分段情况和相应的长度，读取数据包头第一字段的后两个比特判断待发送数据的一个字节是否为一个应用程序的头字节，最后一

个字节是否为一个上层数据的尾字节。

根据对数据包头的分析，解析待发送数据，将解析后的数据进行重组。在实际的应用过程中，可以设置一个缓存，用于暂存完整的上层数据。如果待发送数据的数据域的数据内容只有一段上层数据，且第一字节为一个上层数据的头字节，初始化缓存，将待发送数据的数据内容写入缓存。然后根据待发送数据序列号 SN 是否连续判断是否出现丢包，如果出现丢包，丢弃该待发送数据，同时初始化缓存和其它记录信息，并开始处理下一个待发送数据；如果没有，且最后一个字节为上层数据的尾字节，则该待发送数据的数据内容是一个完整的上层数据，直接交给共享内存，初始化接收信息。如果待发送数据的数据内容存在拼接，获取第一个拼接段，即分段和分段长度并写入缓存，如果分段的第一字节为上层数据的头字节，则该分段是一个完整上层数据，写入共享内存并初始化接收信息；如果不是，是一个上层数据的结尾，并且没有丢包，则缓存得到完整的上层数据，进行处理；然后获取中间的拼接段和分段长度，按照完整的上层数据进行处理；接着获取最后一个拼接段和长度，判断待发送数据的最后一个字节是否为一个上层数据的尾字节，如果是则按照完整上层数据处理；如果不是，则为一个上层数据的开头片段，存入缓存，准备处理下一个待发送数据。

将重组得到的上层数据都存入缓存，然后 PDCP 会将缓存中的上层数据转发给应用层，链路完成。

**步骤 3**，通过 PDCP 层，将上层数据转发至接收端的应用层。到这里，协议结构的过程就实现了从发送端应用层，经过 RLC、MAC 到达 PHY 通过无线发送，以及对应接收端的协议栈功能，对于数据处理过程的结构进行了一定的简化，降低了代码的复杂度，克服了系统稳定性差的缺点，提供了接近理论 LTE 速率的峰值速率，为之后 LWA 系统的设计奠定了基础。

## 3.3 LWA 基站侧的设计与实现

3.1 节从几种不同的 LTE-WiFi 融合方式开始介绍，分别简述了各自在实现架构上的差异以及相应的影响，最后详细的介绍了基于非常紧耦合的 LTE-WiFi 聚合系统的优缺点。LTE-WiFi 聚合系统总体上由发射端的基站侧和接受端的用户侧构成，基站侧是接收应用层的数据，在 PDCP 层进行分流后分别交给 LTE 链路和 WiFi 链路，然后发射出去。在本节中，将对 LTE-WiFi 聚合系统基站侧的设计与实现进行详细地介绍。首先从总体上介绍基站侧的工作流程，然后分模块说明的实现方式。

基站侧的 WiFi 分流主要是在 PDCP 层，在 LTE 中，eNB 传递到 UE 的数据首先要

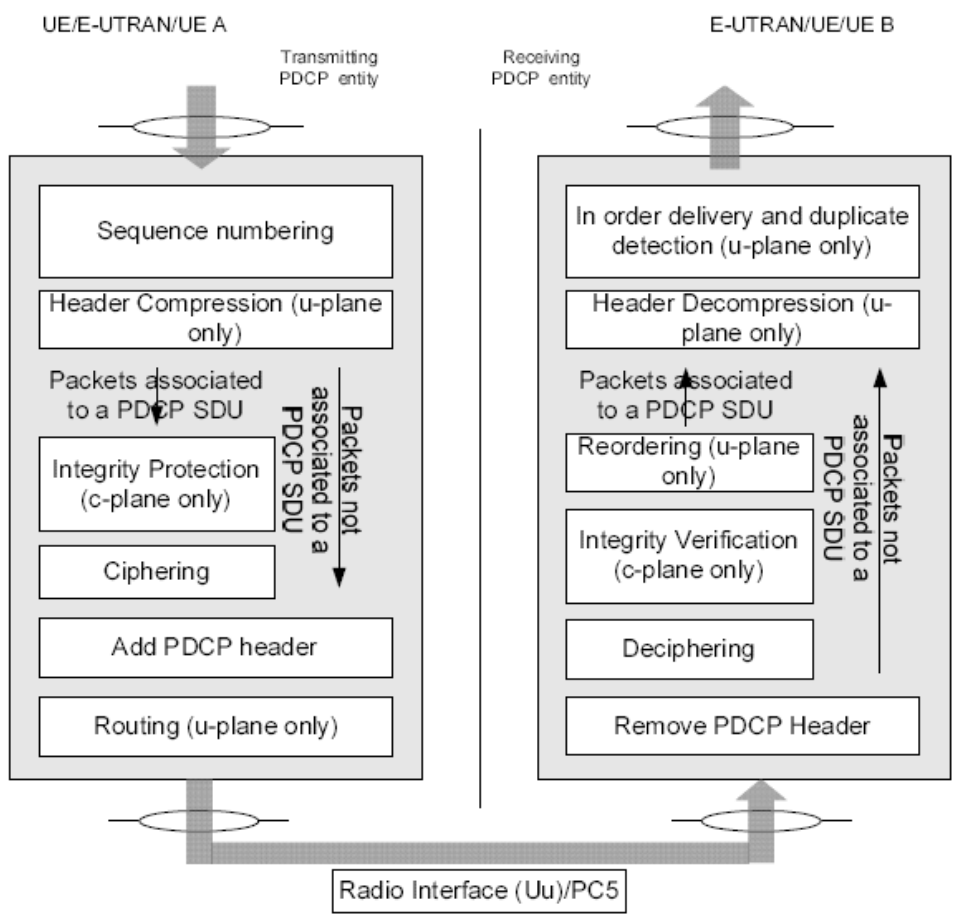

图 3-4 PDCP 层实体结构

经过 PDCP 层然后到达 RLC 层。在 RLC 等待发往外部世界的数据必须要经过 PDCP 层才能到外面的世界。

如图 3-4 所示，进入 PDCP 层的数据首先要编号，也就是 PDCP 要给收到的每一个数据包添加序列号。添加序列号后就需要管理他们。在接收侧，我们就可以借助 PDCP 序列号解决许多事情，比如用户侧的重排序操作。然后，数据进入头压缩过程。但是在标准上这个仅仅适用于用户面数据。也就是说信令信息并不会做头压缩。尽管图中没有给出，我们也可以禁用 U-plane 数据的头压缩（如 ip 包）。

之后，我们可以看到两条路径，一条经过"完整性保护、加密"，另一条直接到下一步。完整性保护仅仅应用于信令信息（即 RRC/NAS 消息， 例如 DCCH 数据，而不是 DTCH 数据）。当然，这里也可用通过设置指令达到禁用"完整性保护"的目的，而且，在实验的初期阶段，可以直接使用另一条透明传输路径。然后是加密过程，控制面和用户面数据都需要加密。加密过程也可以通过指令禁用。不过，在实验的初期阶段，可以直接使用另一条透明传输路径，避免在这一部分出现问题。最终，在 PDCP 传输的最后一步，加上协议头并送出 PDCP 层。

### 3.3.1　基站侧的工作流程

LWA 基站侧由 GPP 和 USRP 两部分组成。本文使用的 GPP 是一台高性能的工作站。GPP 的主要作用是对业务数据进行处理，通过编码调制等流程生成基带信号，然后通过 USB 3.0 接口传输给 USRP，同时将 WiFi 数据包通过 socket 发送。USRP 收到 GPP 传输过来的基带数字信号，通过一系列 FPGA 和数模处理变为射频信号，从天线发射出去。

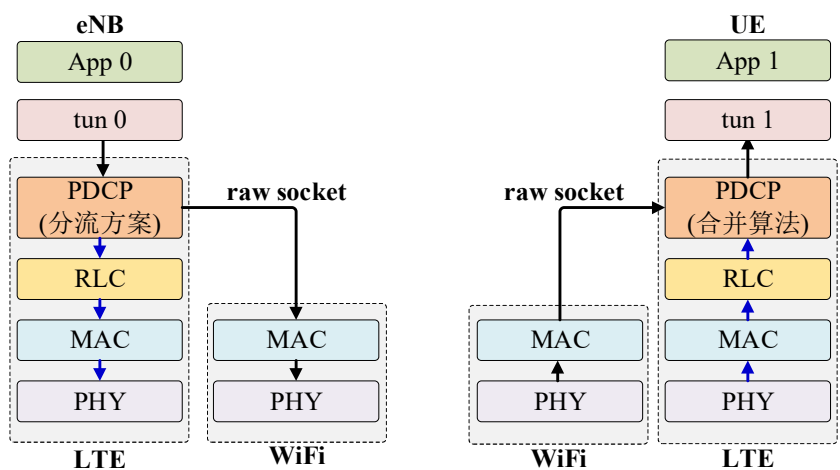

图 3-5 LWA 系统结构图

系统结构如图 3-5 所示，在 eNB 侧，应用层通过虚拟网卡发起流量请求，这里将虚拟网卡称为 tun0。 然后，来自 tun0 的数据被传送到 PDCP 层中进行处理。经过分流操作之后，eNB 将部分数据交给 LTE 网络发送，具体是需要把数据包交给 RLC 层，经过 MAC 层和 PHY 层处理之后，通过通用无线电设备发送；另一部分数据包则需要写成以太帧的格式，使用 Raw Socket 通过 WiFi 网络发送。具体工作流程如下：

在程序开始运行后，首先根据运行时的配置文件和输入参数进行初始化工作。初始化工作主要包括：

● 初始化 log 等级。程序中函数表示为 logInit()，这个函数主要是为了方便开发人员调试程序，在使用过程中，研究者可以根据调试的目的不同设置不同等级的 log 来记录不同的 log 信息，帮助了解运行时的状态，以便于分析问题。

● 初始化物理层参数。

● 初始化 USRP。函数表示为 openair0_device_init()，通过 UHD 连接 USRP 之后，会重置 USRP 设备，包括清除其中的所有缓存等信息，然后将之前所设置的系统参数配置写入 USRP 设备，之后 USRP 进入等待模式，等待程序下一步调度。

● 初始化 L2 层相关数据。程序中涉及到主要的函数有：

■ l2_init()，其作用为初始化 L2 参数；

■ rrc_pdcp_config_asn1_req ()，其作用为初始化 PDCP 层功能，新建 PDCP

entity，初始化基本参数如 pdu 格式，对应的 rlc 模式），当然，在连接转台发生变化时，也可以对 pdcp entity 进行修改或移除；；

- **■** rrc_rlc_config_asn1_req()，其作用为初始化 RRC 层功能，与 PDCP 层初始化类似值得注意的是，rlc entity 会使用哈希表结构保存链表形式的数据包，也是我们分流之后缓存数据的地方。哈希表（Hash table，也叫散列表），是根据关键码值(Key value)而直接进行访问的一种数据结构。也就是说，它通过把关键码值映射到表中一个位置来访问记录，以加快查找的速度。这个映射函数叫做散列函数，存放记录的数组叫做散列表。

- **●** 初始化共享内存以及 socket 通信的相关数据，包括 UE 的物理地址，以太帧结构体等建立连接需要的相关参数。

在完成初始化工作之后，程序通过开始读取虚拟网卡发送到共享内存的数据，并交给 PDCP 层处理，此处是实现 LWA eNB 的关键，将在下一章重点讲述。经过 PDCP 层分流之后，发送数据的操作将在物理层实现。WiFi 数据包通过 socket 发往指定物理地址的用户侧 socket 接收程序。LTE 数据包则在一系列基带处理之后使用 USRP 设备发射。

总体来说，当初始化工作完成之后，LWA 基站就进入等待状态，当检测到虚拟网卡有数据要发送，eNB 就会立即对数据进行处理。当检测到终止程序的命令后，就释放掉占用的系统资源，然后关闭程序。

### 3.3.2 基站侧主要功能模块的设计和实现

为了减少 LTE-WiFi 互通对蜂窝核心网的影响，在 eNB 侧实现了 WiFi 分流功能，其中包括两个主要部分。第一个是关于 LTE 协议的结构。由于在 LTE PDCP 中实现 WiFi 分流，所以有必要对 OAI 系统中的协议实现进行一些改变，主要是 PDCP 层与上层和 RLC 层之间的接口。业务数据到达虚拟网卡后，需要通过 LTE 和 WiFi 发送的数据包将呈现给 PDCP。此外，我们更改了一些基于 OAI 物理层（PHY）的代码，以获得稳定的 LTE 系统。第二个主要问题是分流计划。即分析 LTE 基站负载情况，WLAN AP 负载和链路状态，然后决定是否通过 LTE 或 WiFi 传输数据包到 UE，以充分利用传输能力。关键在于采用什么样的方案可以充分利用 LTE 和 WiFi 网络来提高系统性能[13]。

从基站侧的工作流程可以知道，分流与发送的功能主要需要三个模块，数据读取部分，分流实现模块和 WiFi 数据接口部分。

### 3.3.2.1 数据读取模块

在使用 OAI 的开发时期，我们发现当使用 OAI 自己的网络接口卡时，LTE 速率被限制的情况非常严重。因此，我们采用虚拟网络适配器和共享内存的方式来实现应用层流量的传输。共享内存是共享物理内存的方法，也是在多个进程之间共享数据的快速方式。虚拟网络适配器使用软件来模拟 NIC，这可以是从应用程序接收数据以及通过使用物理适配器将数据通过现有网络传输到接收机的特定 IP 的非常合适的解决方案。添加虚拟网络适配器与应用程序通信，并在应用程序和共享内存之间传递数据，实现虚拟网络适配器和 LTE 设备之间的数据和信令的透明传输。

基站接收到业务请求，上层数据经过虚拟网卡转发放入共享内存，基站将来自共享内存的上层数据存入队列，判断队列内上层数据达到规定较大个数时，取出一个上层数据，并比较上层数据的长度与信道物理资源可传送的数据的长度，如果上层数据的长度大于信道物理资源可传送的数据的长度，则直接进行为上层数据添加数据包头的步骤；如果上层数据的长度小于物理资源可传送的数据的长度，说明需要拼接，则继续接收下一个上层数据，直到总长度大于信道物理资源可传送的数据的长度或者收到的上层数据的个数超过规定可拼接的最大个数，其中，规定可拼接的最大个数在实际的应用过程中预先设定。

所以，程序初始化之后，使用写好的脚本启动虚拟网卡，当应用层业务发起数据时，数据包首先会通过从本地网卡到达虚拟网卡，虚拟网卡把收到的数据包存入共享内存并记录写入的位置等待读取。然后调用函数 read_one_packet_from_shm_near() 从共享内存中取出数据包，判断是否为信令信息，如果不是，则调用 pdcp_data_req()开始处理数据。

### 3.3.2.2 分流实现

当 PDCP 层从共享内存中读取到数据包时，就需要作出判断，下一步是交给 RLC 通过 LTE 发送还是 WiFi 网络。这里我们为了完成整个数据传输的链路，同时又能检测 WiFi 分流的功能，暂时收到的数据包按照固定的比例分给 LTE 和 WiFi。具体操作是在 eNB 全局结构体中设置标识位变量 flag，

```
RLC_UM_MUTEX_LOCK(&rlc_p->lock_input_sdus, ctxt_pP, rlc_p);
rlc_p->buffer_occupancy+=((struct                          rlc_um_tx_sdu_management
*)(sdu_pP->data))->sdu_size;
list_add_tail_eurecom(sdu_pP, &rlc_p->input_sdus);
RLC_UM_MUTEX_UNLOCK(&rlc_p->lock_input_sdus);
```

当 flag 为 0 时，将数据包交给 LTE，当 flag 为 1 时，将数据交给 WiFi。

```
RLC_UM_MUTEX_LOCK(&rlc_p->lock_input_sdus, ctxt_pP, rlc_p);
list_add_tail_eurecom(sdu_pP, &rlc_p->pdus_to_wifi);
```

> RLC_UM_MUTEX_UNLOCK(&rlc_p->lock_input_sdus);

上面的代码使用了一种 pthread 中的锁机制[34]，它们在线程间通信中起到了非常重要的作用。互斥锁是一种常见的信号量，它在本文中被用来防止多个线程[35]在同一时刻访问相同的共享资源。pthread_mutex_lock(&lock)完成了对互斥锁 lock 的锁定，用来保护这个变量。

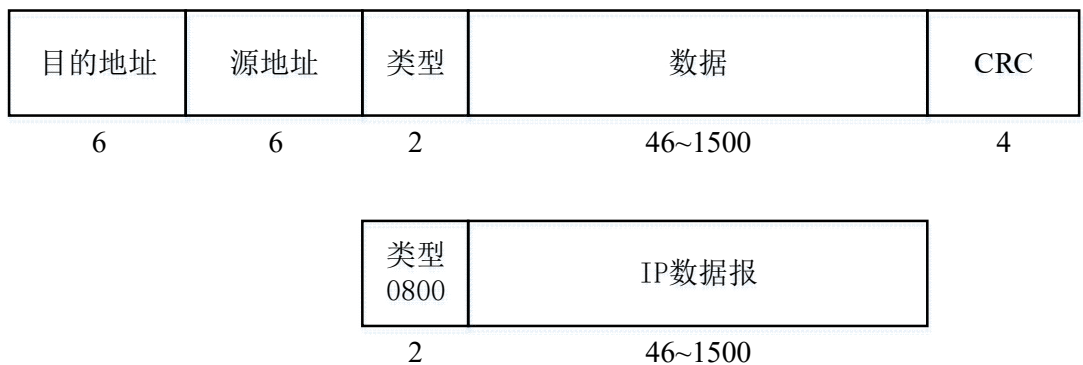

图 3-6 以太帧结构图

### 3.3.2.3 WiFi 数据接口

数据包经过分流之后，会分别放入 LTE 和 WiFi 的处理过程中，LTE 数据会由交给 RLC 层，经过 LTE 链路发送出去，而 WiFi 数据需要添加单独的 WiFi 集成模块，该模块允许 OAI 使用 WiFi 协议在基站与用户侧之间传输数据。这里使用网络套接字（network socket，简称 socket）通信的方式完成 WiFi 数据的发送与接收。Socket 是用于在计算机网络中的单个节点发送或接收数据的内部端点。具体来说，它是网络软件（协议栈）中的这个端点的表示，并且是系统资源的一种形式，使用两个端点用来发送与接收数据时就可以建立一条通信链路。Socket 一般有 TCP 与 UDP 两种通信协议，它们之间的区别主要在于是否基于连接；对系统资源的要求（TCP 较多，UDP 少）；程序实现结构复杂度以及数据的顺序与可靠性，由于系统的处理对实时性要求非常高，所以这里作者选择 UDP 传输的方式，尽量保证发送与接收的操作都不会超时。

Socket 通信另外一个主要的参数是地址，一般多使用 IP 地址，但是由于 IP 地址一般采用 DHCP（Dynamic Host Configuration Protocol，动态主机配置协议）方式获得，会经常发生变化，所以本文中使用一种特殊的 RAW_socket（原始套接字），这种方式可以访问以太帧结构的数据，如图 3-6 所示，使用目的网卡和源网卡的物理地址作为收发地址，这样，即使网络环境发生变化，也不影响 socket 通信。

> data_dl = list_get_head(&rlc_p->pdus_to_wlan);
> **if** (data_dl) {
>     **while** (data_dl){

```
        int     data_size_wlan     =     ((struct     rlc_um_tx_sdu_management     *)
(data_dl->data))->sdu_size;
        memcpy(raw_frame.field.data,          data_dl->data      +      sizeof(struct
rlc_um_data_req_alloc), data_size_wlan);
        raw_frame_len = data_size_wlan + ETH_HLEN;
        if ((raw_ret = sendto(raw_socket_fd, raw_frame.buffer,raw_frame_len, 0, (struct
sockaddr*)&saddrll, sizeof(saddrll))) > 0) {     //成功发送  }
        else printf("***raw_ret = %d***\n", raw_ret);
        data_dl = list_remove_head(&rlc_p->pdus_to_wlan);
        free_mem_block(data_dl);
        data_dl = list_get_head(&rlc_p->pdus_to_wlan);
            }
        }
```

另外，在具体实现时，发送端使用 sendto 函数时，该函数本身只是将数据复制到待发送缓存，几乎没有时延。但是如果没有设置 socket 为非阻塞，在缓存占满之后会进入等待，产生较大延时，从而有可能导致线程处理超时而崩溃。实际上，是因为使用 udp 的 sendto 函数时，会拷贝一次要发送的数据包到传输层的网络缓冲区，然后返回结果。由于这个处理的细节，导致 sendto 函数的返回值并没有反映出真实的网络发送情况。也就是说，从耗时的角度讲，udp 的发送过程基本可以忽略，因为数据拷贝基本不占用多大实际时间。但是对于阻塞的 socket 情况则不同，当网络缓冲区满了以后，sendto 就会阻塞，产生延时。而对于非阻塞的 socket，即使网络缓冲区满了，也会立即返回，不会进行阻塞等待，所以这种情况下正适合于本文的系统发送 WLAN 数据，可以保证传输的实时性，不会引起系统出现超时崩溃的现象。

## 3.4  LWA 用户侧的设计和实现

3.3 小节详细介绍了 LTE-WiFi 聚合系统基站侧的设计和实现，在 eNB PDCP 对应用数据的分流处理，然后分别由 LTE 空口发射和 WiFi 网络传输。在本小节中，将介绍由 USRP 设备收到 LTE 信号后，然后交由用户侧进行处理，解出 LTE 数据，同时接收 WiFi 链路的数据，然后将数据合并，并按应用层的顺序交给上层，完成用户侧的功能。

### 3.4.1  用户侧的工作流程

数据通过空中接口到达 LTE PHY，并通过 WiFi 数据包传送到合并和重新排序功

能。 然后，PDCP 删除 PDCP 头并开始解密。 接下来，对用户数据进行控制数据和报头解压缩检测的完整性验证。 最后，汇聚的数据包传输到上层，以响应请求。

分组的合并和重新排序是在 UE 侧实现的关键过程。 在接收端，UE 经过一系列 LTE 协议栈处理，连同通过 WiFi 传输的数据，将数据包从 RLC 发送到 PDCP。 通过解密等 PDCP 处理，数据包在给予 eNB 之前成为原始数据包，并被传送到上层。 以这种方式，整个链路电路完成并且对应应用层的业务请求实现。 然而，由于传播延迟不同，必须有数据包到达，根据我们的测试台，LTE 的往返时间（RTT）大于 WiFi。 所以在发送到 PDCP 之前，数据需要重新排序，以提高系统的准确性，同时兼顾传输效率。 具体来说，由于 LTE UE 处理时间的限制，重排序功能必须考虑分组到达的每种情况以及时间复杂度。

LWA 用户侧侧的总体工作流程包括以下步骤：

1）初始化工作。与 LWA eNB 侧代码流程相近，当程序开始运行，首先进行初始化工作，根据研究者的设定初始化 log 等级，初始化一些链路信息，初始化物理层参数，初始化 USRP，初始化共享内存、socket 设置等等。这部分工作与基站侧相近。

2）在用户侧的主要处理在函数 phy_procedures_UE_RX() 中完成，主要包括 LTE 方向需要对 RLC 层部分做出修改，将处理后的数据先缓存起来，等待 WiFi 方向从 socket 收到数据后，统一进行重排序操作。

3）对收到的所有数据包遍历重排，将其按照发送的顺序交给 PDCP 层处理。

4）在程序接收到终止命令后，就会终止工作，然后释放程序开辟的内存空间，退出程序。至此，LWA 用户侧程序的工作流程介绍完毕。

### 3.4.2 用户侧主要功能模块的设计和实现

从用户侧的工作流程可以知道，接收侧主要包括以下两个个部分，数据接收，合并与重排序。

#### 3.4.2.1 数据接收模块

数据接收部分分为 LTE 接收和 WiFi 接收部分，其中 LTE 接收在数据从无线接收到 RLC 层不做改动，在 RLC 层处理之后交往 PDCP 层时，需要先将数据缓存到队列中，与 WiFi 数据一起进行处理。WiFi 数据接收部分的实现与基站侧的发送端对应。具体如下

**步骤 1**：根据系统信息构造 hash_key，

```
key = RLC_COLL_KEY_VALUE(ctxt_pP_dl->module_id, ctxt_pP_dl->rnti, ctxt_>pP_dl->enb_flag, rb_idP, srb_flag);
```

**步骤 2**：利用键值获得 rlc 结构体。在此结构体中缓存这 RLC 处理之后的 LTE 数据

链表，之后 WiFi 数据也要缓存在这里。

```
h_rc = hashtable_get(rlc_coll_p, key, (void**)&rlc_union_p);
void *rlc_pP = &rlc_union_p->rlc.um;
rlc_um_entity_t *rlc_p = (rlc_um_entity_t *) rlc_pP;
```

**步骤 3**：使用 socket 接收程序读取 WiFi 数据包，并将数据写入结构体 mem_block_t 中，这是 OAI 提供的数据结构，一种用于生成链表的基本节点。需要注意的是，这里数据段需要去掉以太帧的头信息。

```
read_size = recvfrom(raw_socket_fd, eth_buff, RX_BUFF_SIZE , 0, (struct
sockaddr*)&saddrll, &sock_addrll_len))
eth = (struct ethhdr*)eth_buff;
mem_block_t     *sdu_wlan_p = get_free_mem_block(read_size - 14);
sdu_wlan_p->data_size = read_size-14;
for(int i = 0; i < sdu_wlan_p->data_size; i++) {
    sdu_wlan_p->data[i] = eth_buff[14+i];
}
```

然后把 WiFi 数据链表存入 RLC 结构体中，与基站侧写入数据相同，由于用户侧也有多线程处理结构，所以这里添加链表的时候，也要使用互斥锁保护数据。

```
pthread_mutex_lock(&pdcp_enqueue_lock);
list2_add_tail(sdu_wlan_p, &rlc_p->pdus_to_pdcp);
pthread_mutex_unlock(&pdcp_enqueue_lock);
```

**步骤 4**：WiFi 数据与 LTE 数据合并后，等待下一步处理。用户侧有一个地方不同，因为重排序时可能会对链表中任意位置的数据做处理，所以使用的是双链表，所以在上述代码中函数为 list2_add_tail()。双链表，即双向链表，是一种类似单链表的数据结构，它的每个数据结点中都有两个指针，分别指向直接后继和直接前驱。所以，从双向链表中的任意一个结点开始，都可以很方便地访问和处理它的前后结点。

### 3.4.2.2 用户侧重排序模块

合并后的数据就可以交给 PDCP 层处理，完成传输链路，但是这样存在数据包乱序的问题，由于 LTE 和 WiFi 传输的延时不同会导致按顺序发出的数据包不按顺序到达接收端，还有 WiFi 使用 UDP 方式传输本身也有可能导致数据包不按序到达，所以需要在将数据交给 PDCP 层处理之前做重排序处理，将数据按顺序传给 PDCP 层。

重排序模块要考虑的第一个因素是时间问题，如果处理时间过长会导致系统出现问题，另外，由于数据传输的延时导致接收侧一般会设置接收窗口，超过这个时间到达的数据包会被认为丢失，所以这里重排序也需要综合考虑这些因素，合理设置窗

口，控制等待时间，同时也要及时处理完数据，尽可能高效的将数据包处理并交给 PDCP 层。

该模块的主要流程如下

**步骤 1：** 从结构体中取出合并后的数据链表，读取序号，并保存在一个全局变量 sn 中作为期望收到的序列号。序号使用 PDCP 的 sequence number(SN)，序列号的位数可以自行设置，这里使用 12 bit 的序列号，取值范围是 0~4095。

**步骤 2：** 设置合适的窗口大小 WINDOW_SIZE，然后创建 index，标识当前位置，开始对链表中的数据进行遍历判断，

如果窗口内没有数据包，则跳出循环

读取 index 位置的 sdu（service Data Unit，服务数据单元）的序号 sdu_sn 恰好等于 sn 时，则将数据包交给下一步处理，并将 sn+1，

当 sdu_sn 不等于 sn 时，则需要考虑不同的情况进行相应的处理。如果比 sn 小，则要根据包的序号所处的位置具体分析。

如果出现了 index 等于 window_size 的情况，则说明窗口内没有期望序号的数据包，需要移动窗口

**步骤 3：** 调用 pdcp 层函数 pdcp_data_ind()将取出的数据交给 pdcp 层处理，之后由 OAI LTE 程序完成剩余处理过程，并写入共享内存。用户侧的虚拟网卡就可以拿走业务数据，交给本地网卡，从而完成应用层业务通信的整个链路。

## 3.5  LWA 系统的功能验证

### 3.5.1  系统配置参数

对于我们的实验，如图 3-7 所示，使用具有通用软件无线电外设（USRP）的两个通用计算机。 该计算机的 Intel（R）Core（TM）i7-4770 CPU 具有四个物理内核和 OAI 系统的最大时钟计算要求。 操作系统（OS）在这些计算机中使用 64 位 Ubuntu 14.04。对于 WLAN AP，我们选择市场上常见的无线路由器，其支持通用 WiFi，如 802.11 系列协议。 选择 USRP B210 是因为它提供了一个完全集成的单板 USRP 平台，具有 70 MHz - 6 GHz 的连续频率覆盖。 USRP 专为低成本本实验而设计，结合了完全集成的直接转换收发器，提供高达 56 MHz 的实时带宽，以及快捷方便的 USB 3.0 连接。 完全支持 USRP 硬件驱动（UHD）软件，使我们能够方便地控制和管理 USRP。

表 3-1 实验系统参数配置

| | 类别 | 参数 |
|---|---|---|
| 硬件 | USRP | Ettus Research USRP B210 |
| | CPU | Intel(R) Core(TM) i7-4770 CPU @ 3.40 $GHz$ |
| 软件 | 操作系统 | 64-bit Ubuntu 14.04 LTS with low-latency kernel 3.17.0 |
| | UHD | Version 3.8.0 |
| LWA 系统参数 | 双工模式 | FDD |
| | 下行接入频点 | 2.6 $GHz$ (band7) |
| | CP 类型 | Normal |
| | 传输模式 | TM1(SISO) |
| | 利用子帧数 | 8 |
| | 系统带宽 | 5 $MHz$ |
| | 调制方式 | QPSK,16-QAM,64-QAM |
| WiFi | 协议 | IEEE 802.11n |

### 3.5.2 实验场景

实验场景如图 3-7 所示，我们部署了两台具有 USRP B210 的计算机作为 eNB 和 UE。 UE 连接到连接到 eNB 的 WLAN AP 提供的 WiFi。 主要系统参数如表 3-1 所示。 eNB USRP 的发射功率设置为 10dBm 左右，两个 USRP 在大约两米之外，室内环境无阻塞。 我们使用 iperf 从 eNB 生成 TCP 流量，并测量 UE 的吞吐量。 Iperf 是一种常用的网络测试工具，可以创建传输控制协议（TCP）和用户数据报协议（UDP）数据流，并测量携带它们的网络的吞吐量。

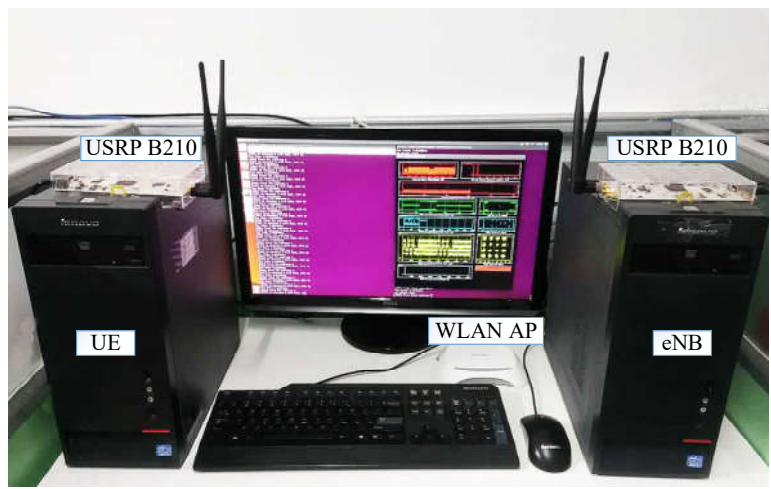

图 3-7 LWA 系统实验环境

### 3.5.3 实验结果及分析

我们知道 LTE 和 WiFi 的延迟是不同的，测试结果列于表 3-2。在电信中，往返时间（RTT）是发送信号所需的时间长度加上时间长度需要接收该信号的确认。因此，该时间延迟包括两个点是 eNB 和 UE 的信号的两个点之间的传播时间。

表 3-2 LTE 与 WiFi 传输时延测试

| RTT(ms) | Min | Avg | Max | Mdev |
|---|---|---|---|---|
| LTE | 3.71 | 5.46 | 7.40 | 1.39 |
| WiFi | 0.86 | 3.20 | 30.32 | 2.66 |

本实验设置 LTE 的带宽是 5 MHz，运行修改添加了 LWA 系统的 OAI 程序，然后把在 LTE 用户侧经过处理后得到的信号 rxdataF_comp 值存入文件，通过 MATLAB 软件对其进行分析，测试程序正常运行时得到的结果如下图所示。可以看到 LTE 链路仍然可以正常运行，星座图都清晰可辨，对于 QPSK、16-QAM、64-QAM 等几种调制方式都没有影响。

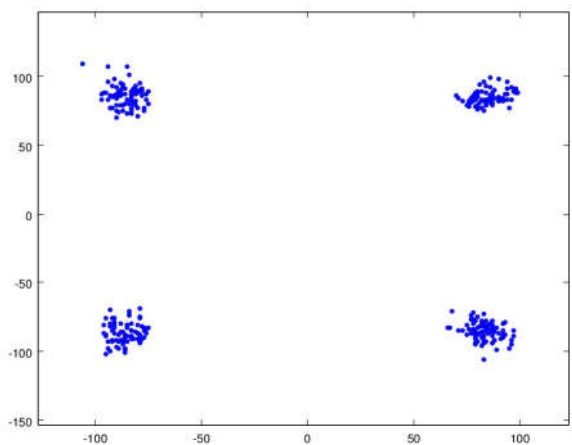

（a）采用 QPSK 调制的信号

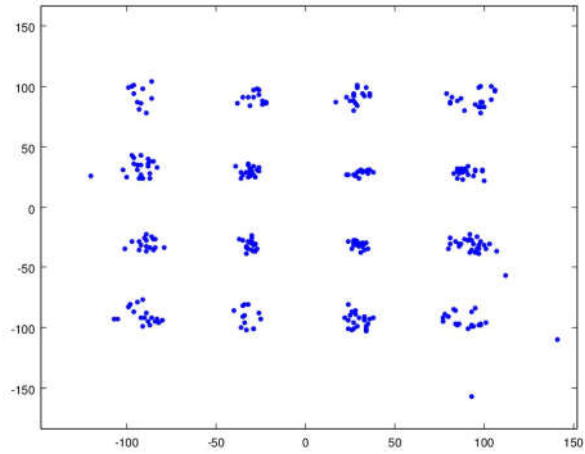

(b)采用 16-QAM 调制的信号

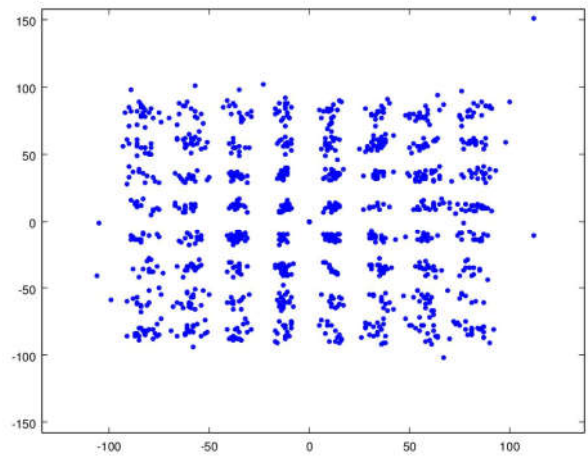

(c) 采用 64-QAM 调制的信号

图 3-8 LTE 不用调制方式下星座图

    首先测试 LTE 与 WiFi 链路各自单独运行的情况，便于分析，在下面的 1）2）中都使用 Iperf 灌包，速率为 20M。

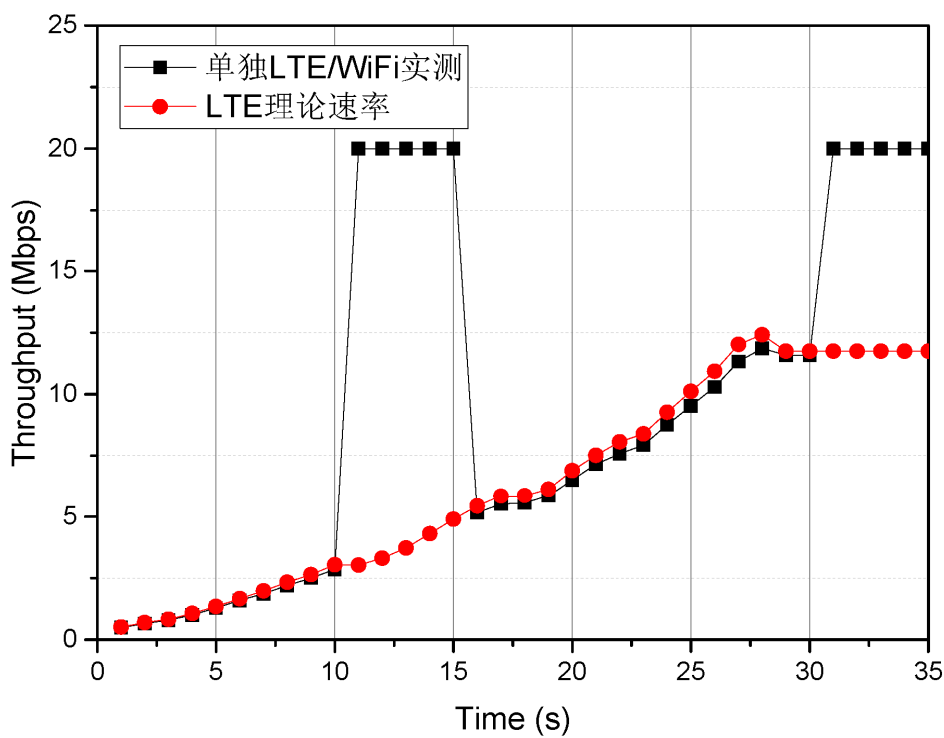

图 3-9 LTE 与 WiFi 测试

1）单独测试 LTE 链路，在这种情况下，数据包完全分流到 LTE，类似于两种技术之间的松散耦合。平均吞吐量的演变如图 3-9 所示的开关模式。应该注意的是，我们在下行 LTE 中只使用 8 个子帧。在全 LTE 情况下，随着高速调制编码方案（MCS）从 1 秒到 27 秒，LTE 的比特率稳定增长，当 MCS 为 27 时达到其门槛值为 28 秒。可以看到，单独使用 LTE 链路时从 0 到 10 秒以及 15 秒到 35 秒，很明显，已经基本达到这种情况下的理论速率。，LTE 吞吐量峰值在大约 11 Mbps，说明系统没有影响 LTE 链路传输能力。

2）单独测试 WiFi 链路：在这种情况下，数据包完全分流到 WiFi，类似于两种技术之间的松散耦合。平均吞吐量的演变如图 3-9 所示。应该注意的是，当 WLAN AP 可用时，整个分流到 WiFi，其中包括两个时间段：当比特率达到约 20 Mbps 时，第一次从 10 秒到 15 秒，第二次从 30 秒到 35 秒。很明显，LTE 吞吐量被限制在大约 11 Mbps，当使用 WiFi 时速率更高而且非常稳定。

3）LTE-WiFi 聚合：在这种情况下，UE 同时使用 WiFi 和 LTE，我们将 LTE 的 MCS 设置为 25.图 3-10 显示了 Ipsf 带宽每 5 秒更改一次的这种情况的比特率。在第一时间从 1 秒到 15 秒，UE 首先开始通过 LTE 发送完整的数据分组，因为 Iperf 带宽低于 LTE 的 Peek 比特率。当 RLC 缓冲器正在被填充时，UE 开始使用越来越多的 WiFi。从 15 秒到 25 秒，LTE 无法承受的其余数据包通过 WiFi 发送，比特率达到 10 Mbps 以上，并且仍在增加。此外，LWA 的比特率几乎是 LTE 和 WiFi 的总和。之后，从 25 秒到 35 秒，当 Iperf 的速率有限时，eNB 再次使用 LTE。根据测试，基于稳定 LTE 的 LWA 系统的

数据分流功能工作良好。考虑到与 WiFi 相比，LTE 的极限速度，将来需要对分裂策略进行测试。

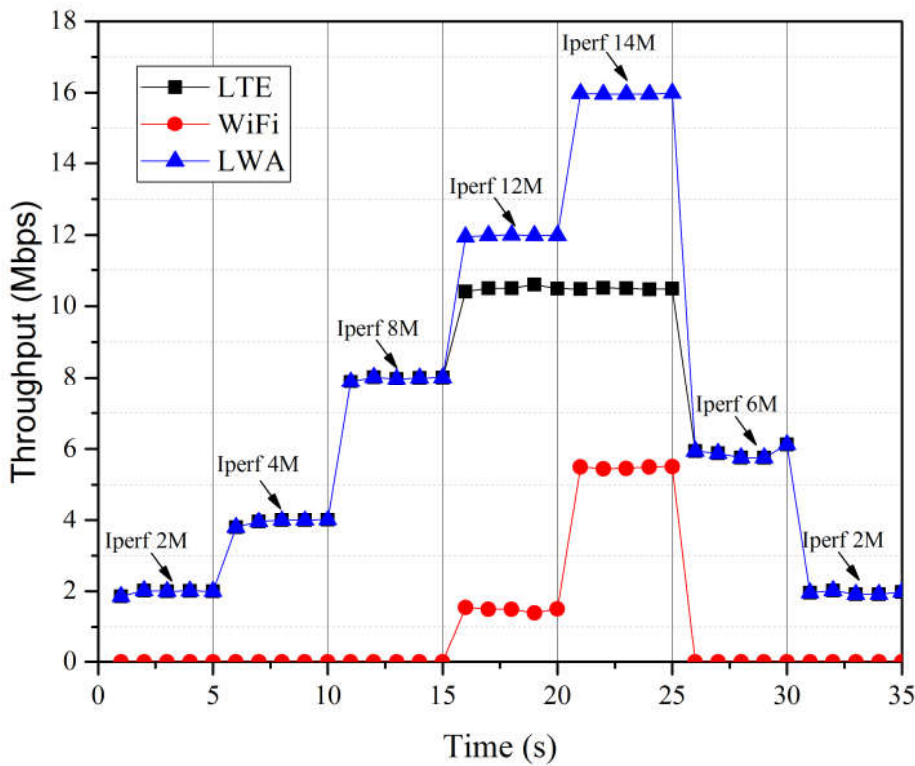

图 3-10 LTE-WiFi 聚合模式

4）重新排序功能：在这种情况下，我们以 LTE-WiFi 聚合方案为例，其中 UE 首先通过 LTE 发送流量，并开始使用 WiFi，而比特率达到 LTE 的峰值。 表 3-3 列出了 Iperf 计数的乱序数据包的比例。 在第一行，我们不使用重新排序功能，直到 Iperf 带宽超过 10 Mbps。 WiFi 开始使用，比特率仍然增加，但随着 WiFi 数据比例的增加，无序乱码的数据包越来越多。 在实验中，使用 PDCP 重排序功能。 无序数据包的数量减少了很多。 但是，由于超时，重新排序功能丢失的数据包更多，丢弃数据包。

表 3-3 实验系统参数配置

| LWA system | 10M | 12M | 14M | 16M |
|---|---|---|---|---|
| Reordering | 0.00 | 0.01 | 0.01 | 0.01 |
| Without reordering | 0.03 | 0.28 | 0.49 | 0.52 |

## 3.6 本章小结

在本文中，我们设计并实现了基于 SDR 的 LWA 系统。 首先介绍了 LWA 技术的原理，并据此对 LWA 基站侧和用户侧进行设计。在开发环境搭建时，着重介绍了对 OAI 协议栈做的优化修改工作。然后详细介绍了收发端的工作流程及重要功能模块的实现方法，主要在于发送端的分流部分和接收端的重排序部分。通过搭建实验环境完成了基于

SDR 的 LTE-WiFi 聚合系统的实时空口通信。实验结果表明，具有 5 M 带宽的 LWA 是稳定和有效的，LTE 链路没有受到影响，基本达到理论峰值速率，WiFi 分流功能也基本实现。然而，本研究实现的 WiFi 分流操作，只是按照预先设置好的比例固定的把数据流量分配给 LTE 和 WiFi 链路，这样明显是没有达到预期的动态调整的分流效果，不能充分利用网络资源。因此，下一章，我们将对于基站侧的分流策略进行优化设计，以提高分流表现性能，使系统为 LWA 的研究提供更多的价值。

# 第四章　LTE-WiFi 聚合系统的分流策略优化设计与实现

在第三章中，本文设计和实现了基于 SDR 的 LTE-WiFi 聚合系统，并在随后测试了系统的基本功能和性能，通过实时空口实验可以看到，数据包可以通过 LTE 链路和 WiFi 链路按照预先设定的比例传输，但是随着数据量的增大，LTE 链路在达到峰值速率后开始出现问题。实际上，LTE 和 WiFi 链路流量的比例可以有一个根据经验制定的初始值，但是，之后需要根据实际网络情况做出动态调整。

上文详细地介绍了 LWA 系统中基站侧和用户侧的工作流程以及主要模块的实现方式，本节的重点是介绍如何在基站侧实现 LWA 数据流的控制模块，主要是根据数据量的不同在 PDCP 层实现数据流的链路选择以及流量分流优化，对于具体的分流策略，我们结合标准的要求以及第三章系统的实现情况，设立了两个 LWA 模式。首先，需要统计在第一种情况下为 LWA 切换模式，当来自应用层的流量低于 eNB 配置的阈值时，系统决定使用 LTE 或 WiFi 网络传输数据。在第二种情况下，如果 PDCP 中可用于传输的数据量超过阈值，系统采用 LWA 分流方式，但是首先，eNB 会根据默认方案在 LTE 和 WiFi 链路上发送流量。阈值的设置主要取决于 LTE 的峰值速率，用来判断是否需要 WiFi 分流。之后，在数据分流后，eNB 根据 LTE 和 WiFi 的链路传输状态进行流量分配，做实时的动态调整，当 LTE 传输受限时分配更多的数据包给 WiFi 链路，反之，当 WiFi 传输受限时，分配更多的数据包给 LTE 链路。

## 4.1　基站侧分流策略工作流程

在 LWA 系统中，由于分流工作主要在基站侧，所以分流策略主要也是在基站侧实现，具体流程如图 4-1 所示，图中有三个变量：$L_i$，$L_{th}$ 和 $L_{re}$，其中 $L_i$ 表示待发送数据量，$L_{th}$ 表示分流门限阈值，$L_{re}$ 表示剩余的数据量。

如前所述，LWA 系统中包含两种传输模式：LWA mode 和 Switch mode。LWA mode 表示系统可以同时使用 LTE 链路与 WLAN 链路传输数据，LTE 链路与 WLAN 链路传输的业务量根据分流比例（m：n）确定。Switch mode 表示系统只能使用 LTE 链路或 WLAN 链路中的一条链路传输数据。系统为两种模式的切换设定了一个阈值，即 $L_{th}$。$L_{th}$ 的值由 LWA Flow Controller 确定。当待发送数据量小于分流阈值时，系统采用 Switch mode，即系统只采用 LTE 或 WLAN 链路中的一条传输数据，具体使用哪条链路根据 LTE 链路与 WLAN 链路的信道条件确定。当待发送数据量大于分流阈值时，系统采用 LWA mode，即系统同时使用 LTE 链路与 WLAN 链路传输数据以提高系统吞吐率，LWA Flow Controller 根据链路状态确定最佳的分流比例。分流阈值与分流比例

可根据系统状态动态调整。

LWA 系统中包含一个控制中枢，即 LWA Flow Controller。LWA Flow Controller 包含三个主要的模块：内容感知模块（Context Awareness Module），链路状态感知模块（Link Status Listening Module）和业务状态感知模块（Data Buffer Listening Module）。内容感知模块根据数据包 TCP/IP 头获取业务内容相关信息；链路状态感知模块根据 LTE 链路与 WLAN 链路的信息反馈获取链路状态；业务状态感知模块则负责实时监控待发送的业务量。LWA Flow Controller 综合三个感知模块的信息制定出最佳的分流策略。LWA Flow Controller 中的三个感知模块相互补充，相辅相成。例如，如果内容感知模块根据协议头中的端口号等信息判断业务类型为 FTP 业务，则应保障业务传输的可靠性；如果感知其为流媒体数据，业务的实时性比可靠性更重要，则应为业务分配更高的传输优先级。同时，LWA Flow Controller 需要实时感知系统待发送的业务量，如果待发送业务量增大，系统应为信道条件较好的链路分配更多的数据以提高系统的吞吐率。链路状态感知模块是 LWA Flow Controller 的基础，LWA Flow Controller 只有在正确、充分获取链路状态信息的基础上才能根据业务内容和业务量做出最佳的分流策略。

系统运行流程如下：

- 系统启动，各模块初始化
- LWA Flow Controller 更新业务内容、链路状态信息和业务量等信息
- 系统从 LWA Flow Controller 中获取 Li，如果系统无数据待发送，则短暂休眠
- 如果系统有数据待发送，系统从 LWA Flow Controller 中获取 Lth。如果 Li < Lth，系统采用 Switch mode，即使用一条链路传输数据即可；如果 Li >= Lth，系统采用 LWA mode，即同时使用 LTE 链路与 WLAN 链路传输数据。具体的分流策略由 LWA Flow Controller 综合系统信息确定
- 本次传输完成后，系统根据 Lre 与 LWA Flow Controller 中信息更新 Li

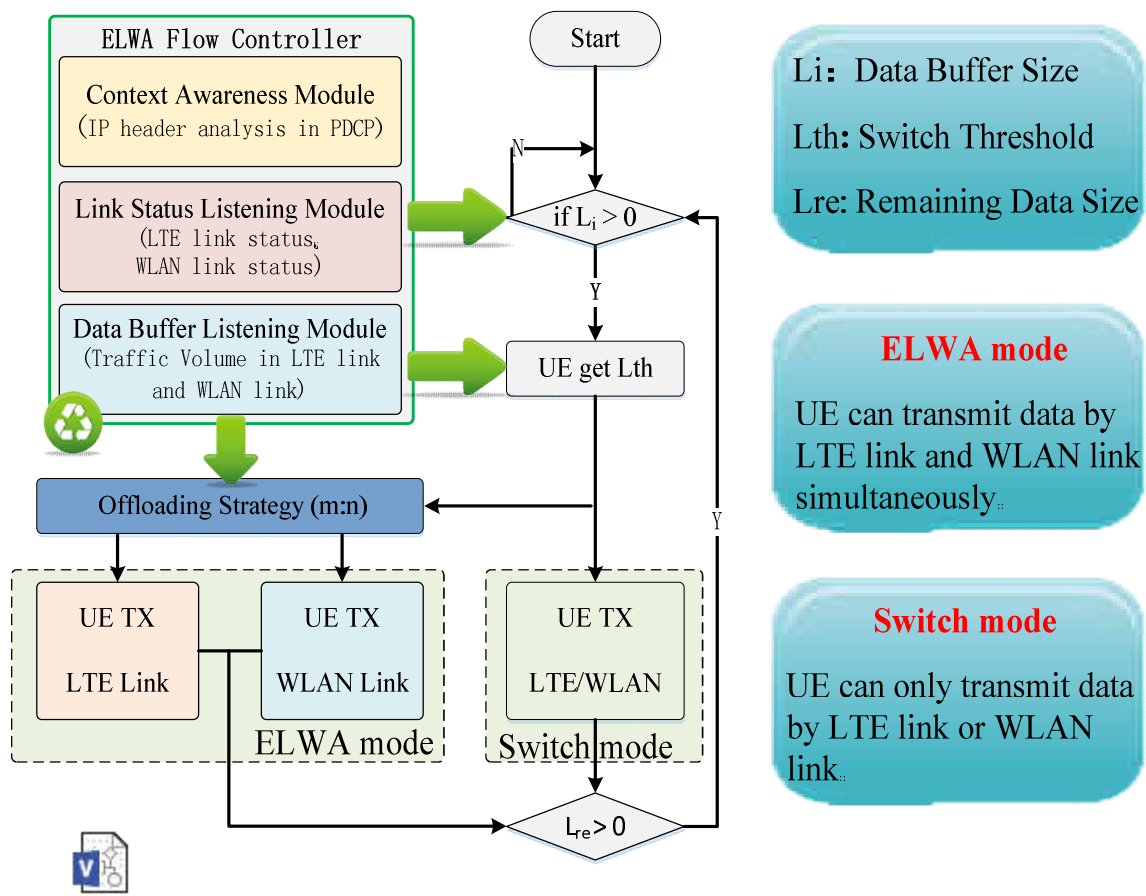

图 4-1 LWA 系统的分流策略优化

## 4.2 主要功能模块设计与实现

但是，本文实现的 LWA 系统重点是在下行链路，由于没有完整的上行链路，所以有些需要完整回路的业务暂时无法运行，测试使用的都是单向的 Iperf 业务，所以，在控制中枢中，内容感知模块目前并没有实现。所以，可以得出，基站侧的分流控制模块主要分为以下三个部分，首先是门限判别，在基站侧设定阈值，并统计来自应用层的数据量，与之比较。然后是链路状态感知，对 LTE 和 WiFi 链路传输情况做实时监控。最后是分流策略更新，完成对数据包的流量分配，通过 WiFi 和 LTE 空口传输。

### 4.2.1 门限判别

在做分流阈值的门限判别之前，首先要计算待发送数据量。待发送数据量用来表示当前业务量的大小，也就是一段时间内，基站侧需要发送的数据包的总长度。本文从方便统计的角度出发，在 FDD 的帧结构中，如图 4-2 所示，1 帧为 10ms，分为 10 个子帧，100 帧为 1s，所以选取 100 帧为时间单位，计算这段时间内要发送的数据包的总长度。

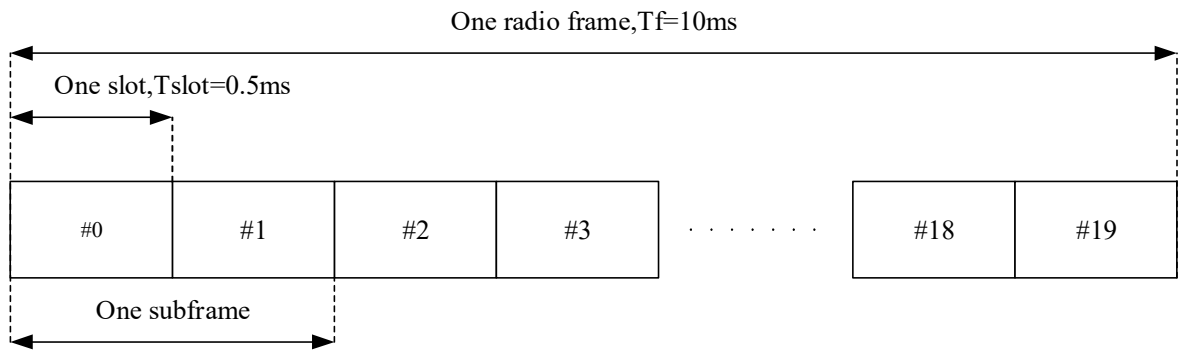

图 4-2 LTE 物理层帧结构

　　而需要统计的数据包，则是 PDCP 层从虚拟网卡 tun0 读取到的。这里，由于需要统计每一帧的数据包，所以在实现时，在 eNB 的全局的结构体中新建了变量 Li，累加 100 帧内 PDCP 层收到的数据包长度。之后，就是与判断是否分流的门限阈值进行比较。阈值一般设置为 LTE 的峰值速率，不过可以根据需要和实际情况乘以一个门限因子，比如为了测试之后的链路状态感知，不希望开始分流的时候 LTE 链路就已经饱和，而是留有一定的余量。

　　在得到待发送数据量 Li 和分流门限阈值 Lth 之后，只需要对 Li 和 Lth 做一个比较，如果 Li < Lth，系统采用 Switch mode，即使用一条链路传输数据即可；如果 Li >= Lth，系统采用 LWA mode，即同时使用 LTE 链路与 WLAN 链路传输数据。

### 4.2.2　链路状态感知

　　在门限判别之后，LWA 控制中枢会选择一种传输模式，当数据量较低时，LWA 处于切换模式，使用 LTE 链路或者 WiFi 链路传输，这时链路状态感知模块并不需要工作。只有当数据量增大，超过门限阈值时，LWA 进入分流模式，才开始对链路状态进行监控。

　　由于缺乏上行的反馈信息，无法通过用户侧或者核心网来获得 LTE 或者 WiFi 的链路状态，所以这部分的感知工作主要是在基站侧进行。在分流实现模块，首先需要申

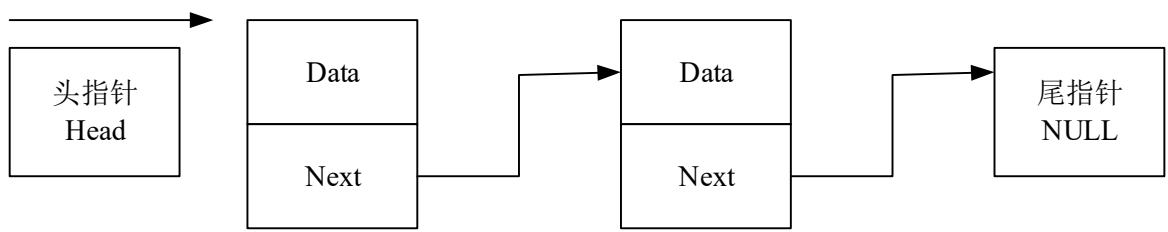

图 4-3 单链表结构图

请缓存，分别命名为 lte_data_list 和 wifi_data_list，用来保存 LTE 链路待发送的数据包和 WiFi 链路待发送的数据包。这里的 list 指单向链表。单向链表（单链表）是链表的一种，其特点是链表的链接方向是单向的，对链表的访问要通过顺序读取从头部开始；

链表是使用指针进行构造的列表；又称为结点列表，因为链表是由一个个结点组装起来的；其中每个结点都有指针成员变量指向列表中的下一个结点，这样可以方便的对不同链路的数据包按序进行存取。而当 LWA 进入分流模式时，首先会采取初始化的比例（默认为 1:1），把数据包分别存入两个链表，然后 LTE 链表中的数据交给 RLC 层，通过空口发送，WiFi 链表中的数据则是通过套接字发送，这时，才开始进行链路状态感知，具体的做法是：

首先，记录当前时刻两个链表的长度，然后，每隔一段时间（这里需要根据测试情况进行调整，本文目前以 10 帧为单位），再记录两个链表的长度，计算链表的长度变化。当链表长度没有增加时，说明对应的发送链路的输送能力可以满足目前分到的数据量。反之，当链表长度开始增加时，即对应的发送链路中出现了待发送数据包积压的情况，说明当前链路已经饱和，需要调整分流比例，将更多的数据包分给另一链路。

具体在调整分流比例时，最重要的是确定调整幅度的大小。如果幅度太小，则饱和链路积压情况得不到缓解，无法起到动态调整的作用。而幅度太大，则可能产生新的问题。所以，本文在设置调整幅度时，也采取了动态调整的方式：每次得到计算的链表数据包积压增量时，用其除以一个基准量，得到调整幅度，这样，当一个链路的数据量突然增大时，可以大幅调整分流比例，而当数据量的积压情况缓解时，仍然能够以较小的幅度调整分流比例，起到更好的分流效果。

### 4.2.3　分流策略更新

如前所述，分配给 LTE 和 WiFi 的数据分组缓存在两个列表中，分别等待发送。经过一段时间，由于 LTE 和 WiFi 链路状态和业务量的变化，不同的传输容量可能导致链表中的数据发送积压。此时，eNB 根据两个列表的长度修改分发策略，将更多的分组分发到 LTE 或 WiFi。

在 4.2.2 中，根据链表状态变化确定了分流的调整幅度，这时，只需要根据调整幅度计算新的分流比例即可。

首先，需要判断链路状态，如果两个链表都没有发生积压，则不需要更新分流比例。当一方发生数据包积压时，将默认值乘以调整幅度，得到该链路新的分流比重。但是由于实际在分流时以数据包为单位，所以，还需要对刚刚计算的分流比例进行一定的处理，使得 PDCP 层可以在单位时间内将对应比例的数据包写入不用的链表中。通过连续监控两个列表的数据包，eNB 就可以在较短时间内（本文中为 10 帧，即 0.1s）更新两种网络的分流比，充分利用两种技术尽可能多地传输数据。

## 4.3 算法性能验证

### 4.3.1 系统参数配置

本节的系统参数配置与 3.5.1 中系统配置相同，此处不再赘述。

### 4.3.2 实验场景

本节的实验场景与 3.5.2 中实验场景布置相同，此处不再赘述。

### 4.3.3 实验结果及分析

以下测试门限因子为 0.8，门限采用 LTE 在 5 MHz 带宽下的峰值速率，可以计算得到，此时理论峰值速率(8 个子帧，约 14 Mbps)乘以门限因子，约为 11 Mbps。另外，为了更加直观的对 LTE 和 WiFi 各自的流量进行观察，这部分测试将结果做了可视化处理。利用 Octave 和 Gnuplot 工具，将代码中统计的速率通过画折线图的形式显示出来，可以清楚的看到两条链路的速率变化。Octave 是一种科学计算软件，简单来说，可以将其认出是 Linux 环境下的 Matlab，与之非常相似而且语法高度兼容。Octave 提供了方便的互动命令列接口来解决线性与非线性的数值运算问题，并可将计算结果可视化。而 Gnuplot 是一个命令行的交互式绘图工具。用户通过输入命令，可以逐步设置或修改绘图环境，并以图形描述数据或函数，使我们可以借由图形做更进一步的分析。它支持二维和三维图形。它的功能是把数据资料和数学函数转换为容易观察的平面或立体的图形，它有两种工作方式，交互式方式和批处理方式，它可以让使用者很容易地读入外部的数据结果，在屏幕上显示图形，并且可以选择和修改图形的画法，明显地表现出数据的特性。

1）LTE-WiFi 聚合：如图 4-4 所示，在这种情况下，测试 Iperf 灌包速率从低到高时，LWA 系统对应不同流量所做的分流策略。可以看到，开始测试时，速率较低，可以看到，此时速率低于触发分流模式的门限，系统根据默认的设置将数据包全部交给 LTE，此时 WiFi 没有速率。当 Iperf 速率不断增加，超过 11 Mbps 时，系统根据计算判定开始进入分流模式，此时开始将部分数据包交给 WiFi 发送。接下来，发送端速率不断提高，可以看到，LTE 速率已经已经达到设定的最大值，此时，LWA 将更多的数据包交给 WiFi 链路，并且根据两条链路状况，会进行一定程度的调整。

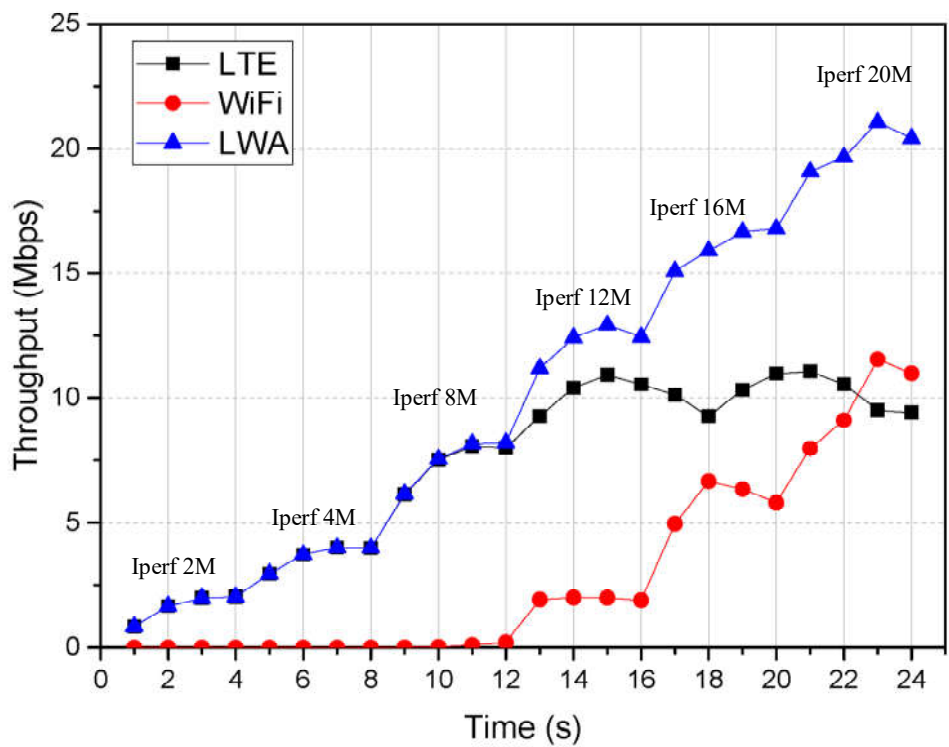

图 4-4 LTE-WiFi 聚合 UE 速率图

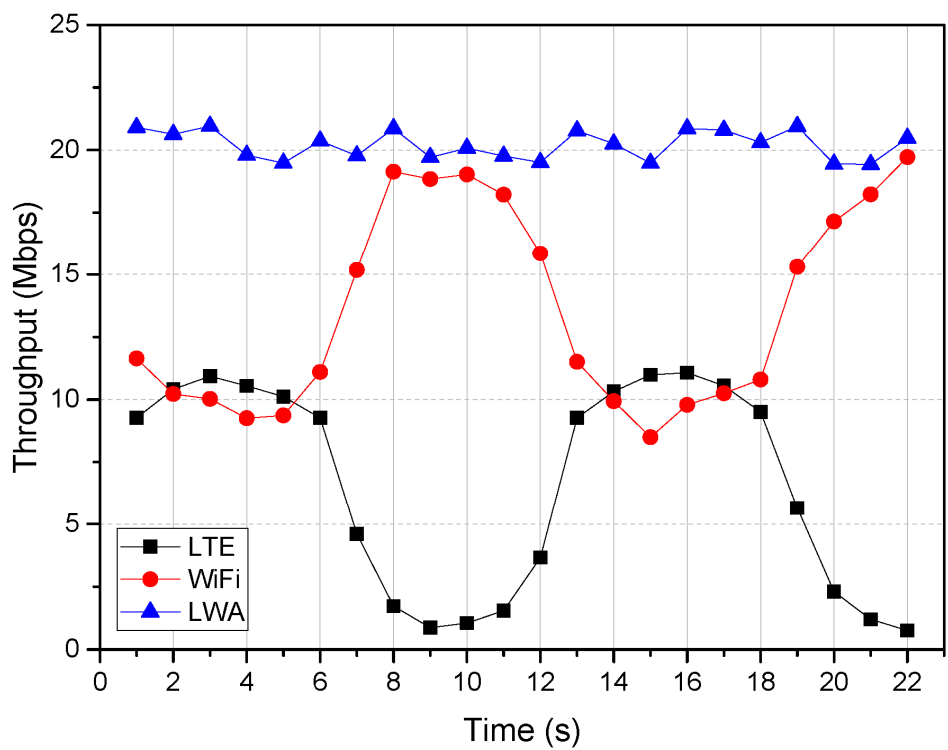

图 4-5 LTE 受限时 UE 速率

　　2）LTE 受限：如图 4-5 所示，在固定灌包速率下，等待 LTE 与 WiFi 链路都达到期望速率后，在一段时间内，使用工具遮挡 LTE 传输，使得 LTE 信道条件变差，从而导致其速率下降，在图中可以看到两次这样的实验，分别是从 6 秒到 13 秒以及 18 秒之后，此时 UE 测得的总速率几乎不变，其中，WiFi 速率逐渐增加，补偿了 LTE 因为信

道受限缺少的传输能力，由此可以看出，LWA 系统根据链路实时的状态进行了动态调整，将数据包更多的交给信道条件、传输能力更好的一方。

## 4.4 本章小结

  本章主要介绍了 LTE-WiFi 聚合系统分流策略优化。系统在数据量与信道环境发生变化时，通过对数据缓存的记录整理，进而实时分析链路状态，达到动态调整的目的。本章首先介绍了基站侧分流策略的工作流程。然后对分流策略的三个功能模块分别作了详细的介绍，包括系统模式的切换以及如何判别网络传输状态并进行动态调整。并在随后针对分流策略设计了不同的测试场景进行测试，测试结果证明系统可以根据流量进行模式切换，验证了动态调整功能可以实现，并取得了不错的分流效果。虽然在应用层业务等方面仍然有待改善，但是，由于设计充分利用了平台的优势，所以系统还具有很强的灵活性和扩展性，能够为新技术的验证提供可靠的参考。

# 第五章　总结与展望

随着移动互联网时代的到来，移动流量的爆炸式增长对蜂窝网络的承载能力提出更高的要求。在这样的背景下，当蜂窝网络的传输能力无法再短时间内获得大幅提升时，WiFi 网络作为一种重要的补充技术对于分担负载压力有着先天的优势，因此当前情况下对于 WiFi 分流技术的研究显得尤为迫切。LTE-WiFi 聚合技术作为一项被人们寄予厚望的新技术，其功能实现和性能的验证研究工作也非常重要。从频谱上来看，随着蜂窝网变小，WiFi 生态系统急剧扩张，蜂窝和 WiFi 都需要额外的频谱。随着 LTE-WiFi 聚合解决方案的发展受到 3GPP 和 IETF 的推动，关于共享接入频谱的更多讨论也将很快展开。而且，随着硬件技术的发展带来的通用计算机性能的不断提升，SDR 平台由于具有灵活性高，扩展性强、成本低廉等优势获得了广泛关注，因此本文设计并实现的基于 SDR 的 LTE-WiFi 聚合系统，对于研究者验证 LTE-WiFi 聚合技术甚至更多 WiFi 分流的新技术都十分有意义。

本文首先介绍 LTE-WiFi 聚合技术提出的背景，并对 WiFi 分流技术的研究现状进行了总结。然后对 SDR 技术的发展和现状进行了简要的介绍，最后给出本论文的文章结构安排。

在第二章中，通过对国际标准组织多年来工作的总结，对 WiFi 融合系统的研究和发展状况做了简要概述，之后对本文使用到的软硬件平台进行了详细介绍。软件部分对 SDR 技术进行了深入的研究，对在 SDR 平台上发展起来的几款开源软件平台做了对比，通过对其功能、稳定性、可扩展性以及影响力的分析，本文选择了 OAI 平台作为实现基于 SDR 的 LTE-WiFi 聚合系统的基础。然后对实验中使用到的通用硬件设备 USRP 等进行了简单的介绍。

第三章主要介绍了基于 SDR 的 LTE-WiFi 聚合系统的设计与实现。包括在数据读取部分虚拟网卡和共享内存的添加，WiFi 集成模块的实现，以及用户侧对 LTE 链路的 WiFi 链路数据的合并和重排序，不过在这一章中，为了验证系统的链路完整性，分流操作并没有考虑实际的网络传输状况，只是将数据流按照固定的比例进行分配，本文第四章将提出一种分流策略的优化设计，从而解决这个问题。

第四章在 LWA 系统的控制中枢部分提出了分流策略的优化方法和实现。首先，主要是依赖于分别为 lte 和 wifi 数据包建立的缓存链表。通过对两个链表的状态分析，推测出 lte 和 WiFi 网络的传输情况，并据此对分流比例进行动态的、实时的调整，从而达到充分利用两种无线网络的目的。

当然，由于作者的水平和研究时间的限制，本文所提出的 LWA 系统平台还存在一

些不足之处，在下一步的工作中可以进行完善和补充，主要有以下两方面可以改进：

- 本文的实验场景均为室内环境，没有其它干扰信号，信道状况理想，所以本文所得到的数据都非常接近理论值。另外在测试链路受限的情况时，在实验过程中通过遮挡直射径可以容易的模拟 LTE 传输受限的情况，但是由于 LTE 系统峰值速率的限制，对于 WiFi 受限场景的测试没有取得理想的效果，这一方面的测试有待改善。如何提升 LWA 系统的性能和稳定性，这是后续工作的重点。

- 本文所设计的分流策略都是在基站侧进行，主要也是利用基站侧的数据分析实现。由于受到上行链路的影响导致测试以单一业务为主，所以在测试内容和数据分析的广泛性上也有所不足，而且因为这个原因，LWA 控制中枢的内容感知模块暂时也没有实现，这一部分仍然需要做更多的工作，使平台性能更加稳定，功能更加完善，这也是作者希望未来能够达成的目标。

# 参考文献

# 缩略语

| LWA | Non-orthogonal Multiple Access | LTE-WiFi 聚合 |
|---|---|---|
| SDR | Software Defined Radio | 软件定义无线电 |
| GPP | General Purpose Processor | 通用处理器 |
| USRP | Universal Software Radio Peripheral | 通用软件无线电外设 |
| LTE | Long Term Evolution | 长期演进 |
| DCI | Downlink Control Information | 下行控制信息 |
| DSP | Digital Signal Processor | 数字信号处理器 |
| FPGA | Field Grogrammable Gate Array | 现场可编程门阵列 |
| FDD | Frequency Division Duplexing | 频分双工 |
| 5G | the Fifth Generation | 第五代通信系统 |
| TDD | Time Division Duplexing | 时分双工 |
| 3GPP | the 3$^{rd}$ Generation Partner Project | 第三代合作伙伴计划 |
| UE | User | 用户 |
| eNB | evolved NodeB | 基站 |
| EPC | Evolved Packet Core | 核心网 |
| UHD | USRP Hardware Driver | USRP 硬件驱动 |
| ADC | Analog-to-Digital converter | 模数变换器 |
| DAC | Digital-to-Analog converter | 数模变换器 |
| CPU | Central Processing Unit | 中央处理器 |
| MCS | Modulation and Coding Scheme | 调制与编码策略 |
| DLSCH | Downlink Shared Channel | 下行共享信道 |
| ULSCH | Uplink Shared Channel | 上行共享信道 |
| SDU | Service Data Unit | 业务数据单元 |
| RB | Resource Block | 资源块 |
| RE | Resource Element | 资源元素 |
| PDU | Protocol Data Unit | 协议数据单元 |
| PDCP | Packet Data Convergence Protocol | 分组数据汇聚协议 |
| RLC | Radio Link Control | 无线链路层控制协议 |

| PDCCH | Physical Downlink Control Channel | 物理下行控制信道 |
|-------|-----------------------------------|------------------|
| QPSK | Quadrature Phase Shift Keying | 正交相移键控 |
| QAM | Quadrature Amplitude Modulation | 正交幅度调制 |

# 致　谢

两年半的研究生时光转眼逝去，现在还能想起刚进实验室的困惑和迷茫。两年半的时间让我逐渐成长，从刚开始时面对问题紧张、不知所措，到现在有了一套自己思考问题、分析问题、解决问题的思维方法，在实验室的这段时间我学会了很多很多。当论文写到这里，我的学生生涯也即将画上句号，新的职业生涯也将拉开帷幕。在这

首先，要感谢我的导师王健全教授。王老师不仅在学术研究上给我了很大的帮助，还能在我陷入迷茫时解疑答惑，是我人生路上的一位重要导师。王老师渊博的学识，对待学生认真严谨的态度，以及他平易近人的人格魅力都对我产生了深远的影响。我在研究生阶段取得所有成果都离不开王老师的指导和帮助。对于我毕业论文的指导，王老师更是尽心尽力。从选题到具体内容的安排再到最终论文的完成，每一步都倾注了王老师大量的心血。虽然我的研究生生涯即将结束，但是王老师将会是我一直学习的榜样。

其次，我要感谢王文博教授。在王老师的带领下，WSPN 实验室学术氛围浓厚，整个团队团结友爱，积极向上。实验室老师们认真负责，学生们充满活力，能够在这样一个实验室中学习和成长，我感到非常幸运。

同时，我也衷心的感谢郑侃教授和赵慧副教授。郑老师特别注重对学生综合能力的培养。赵老师为我们组织了职业技能培训，给我们开辟一个交流平台让我们能够锻炼自己的演讲能力，还会给我们安排各式各样的素质拓展活动。正是因为赵老师的细心指导，我在实验室中不仅提高了自己的研究能力，还全面提高了自己的综合素质。郑老师有着非常扎实的理论基础，每当遇到困惑时，郑老师总能给我解疑答惑，同时郑老师经常与学生一起参加体育活动，对我来说，郑老师亦师亦友，祝郑老师在以后工作生活中一切顺利。

我还要感谢实验室里的所有人。因为你们让我的研究生生活丰富多彩。感谢熊雄博士对我的耐心指导，以及项目团队中申衡阳师兄、魏兴光、耿志明，是团队的共同努力才让我能够顺利完成任务。还要感谢宿舍的好兄弟王浩楠、杨孝康、余雷，正是他们在生活和学习上对我的帮助和鼓励，让我顺利度过了研究生生涯。

最后我还要感谢我的父母。是他们含辛茹苦地把我养大，是他们一直以来对我不计回报地付出，让我能够走到今天，他们是我不断拼搏、不断向前的最大动力。

由于我的学术水平有限，论文难免有不足之处，恳请各位老师批评和指正！

# 攻读学位期间发表或已录用的学术论文